# Дырочная проводимость в неоднородных фрагментах ДНК

**Пономарев О.А.[*1], Шигаев А.С.[**1], Жуков А.И.[2], Лахно В.Д.[1]**

[1]*Институт математических проблем биологии, Российская академия наук, Пущино, Московская область, 142290, Россия*

[2]*Московский государственный университет дизайна и технологий, Москва, 115035, Россия*

***Аннотация.*** Особенности миграции катион-радикала (дырки) в гетерогенной ДНК исследованы с помощью замкнутой системы уравнений корреляционных функций для коэффициента проводимости на основе выражения Кубо. Замыкание цепочки производилось путем отбрасывания корреляций выше второго порядка. Полученная система нелинейных дифференциальных уравнений была исследована как аналитически, так и численно. С помощью формулы Кубо изучена относительная подвижность дырки в гетерогенных фрагментах ДНК. Продемонстрирована важная роль поляризации среды, обусловленной взаимодействием дырок с колебаниями нуклеотидных пар, в процессах переноса. Энергия данного взаимодействия может превышать энергию колебаний в 10 раз. Показано, что скорость переноса между соседними парами оснований ДНК в одномерном случае почти не зависит от более удалённых пар. Осцилляции плотности заряда происходят в пикосекундном масштабе времени, однако скорости переноса дырки, получаемые усреднением по этим осцилляциям, оказались весьма близкими к экспериментальным данным. Полученная зависимость скорости переноса дырки между двумя гуаниновыми основаниями от числа находящихся между ними адениновых оснований также хорошо согласовалась с экспериментами. Кроме того, исследована зависимость скорости переноса от температуры. Показано, что основной вклад в процесс транспорта заряда при 300 К вносит прыжковый механизм переноса.

## ВВЕДЕНИЕ

Наличие в молекуле ДНК системы ароматических гетероциклов, плотно прилегающих друг к другу, обусловливает её проводящие свойства. В силу своей способности к самосборке ДНК является перспективным материалом для создания компьютеров нового поколения. Поэтому в последние годы проводящие свойства этой молекулы вызывают всё больший интерес [1–4].

Проводимость ДНК была впервые исследована ещё в начале 1960-х годов, вскоре после открытия Уотсоном и Криком спиральной структуры дуплекса [5]. Впоследствии показано, что перенос катион-радикала (дырки) в дуплексе не только играет важную роль в процессах канцерогенеза и мутагенеза [6, 7], но участвует также в устранении повреждений ДНК [8–10].

[*] olegpon36@mail.ru
[**] shials@rambler.ru

Основным источником дырок в ДНК являются радикалы гидроксила $OH^{\bullet}$ и другие свободные радикалы, возникающие при воздействии ионизирующих излучений и атакующие дуплекс из раствора [11]. Дырки, мигрирующие в ДНК, локализуются преимущественно на гуаниновых основаниях, *dG* [7, 12, 13] Здесь и далее мы опускаем приставку «дезокси-», обозначая её лишь в сокращении. Локализация катион-радикалов обусловливает реакции радикализированных оснований с водой и кислородом, что приводит впоследствии к точечным мутациям [14] или разрывам цепи дуплекса [15].

Если перенос дырки на *dG* невозможен, происходит окисление тиминовых оснований, *dT* [16], хотя их потенциал окисления несколько выше, чем у комплементарных им адениновых оснований, *dA* [17, 18]. Большие группы смежных *dT* обычно находятся в областях ДНК, не несущих генетической информации и их окисление менее опасно для живой клетки. Поэтому экспериментальные и теоретические исследования закономерностей переноса заряда в ДНК крайне актуальны для радиобиологии и медицины. Открытие эффективного механизма управления скоростью переноса катион-радикала в ДНК откроет большие перспективы, например, в терапии раковых заболеваний. Кроме того, подобные исследования крайне актуальны для развития такой относительно новой науки как нанобиоэлектроника [1, 3].

Исследования проводящих свойств ДНК весьма многочисленны и разнообразны, а их результаты пока трудно привести в систему. По одним данным, ДНК может вести себя как полупроводник [19] или даже изолятор [20]. По другим, она способна быть хорошим проводником [21], а при определённых условиях – даже сверхпроводником [22]. Переход полимеров в высокопроводящее состояние хорошо изучен, например, для полиаренфталидов, в которых он происходит в результате таутомерного перехода, идущего по триггерному механизму [23]. Возможно, что ДНК в этом плане в какой-то мере аналогична полиаренфталидам.

Главным фактором, определяющим проводимость ДНК, является её нуклеотидная последовательность, поскольку азотистые основания, входящие в состав нуклеиновых кислот, обладают разными потенциалами ионизации. Показано, что последовательности ДНК вида поли(*dA*):поли(*dT*) могут вести себя как полупроводники n-типа, тогда как гуанин-цитозиновые гомополимеры обладают проводимостью p-типа [24]. Возможность выпрямляющего эффекта в несимметричных гетерополимерных ДНК продемонстрирована в теоретических исследованиях Шмидта с соавт. [25].

Прямые измерения вольтамперных характеристик ДНК требуют соблюдения ряда условий. Прежде всего, нужно, чтобы был исключён перенос заряда между электродами вне ДНК. Обычно это достигается нанесением амфифильного вещества на участки поверхности электродов, не контактирующие с концами дуплексов [26, 27]. Кроме того, необходима надёжная ковалентная связь между дуплексом и электродами: без неё монослой ДНК превращается в изолятор [28]. Но, несмотря на технические сложности, эксперименты этого типа дают ценную информацию о свойствах ДНК.

Помимо прямых измерений, проводящие свойства дуплексов изучают с помощью физико-химических методов, в ходе которых ДНК приобретает избыточный заряд в результате химических реакций. Исследования этого типа позволяют изучать миграцию заряда на сколь угодно малые расстояния. Скорость переноса между донором и акцептором, $k_{\mathrm{ET}}$, для небольшого гомогенного фрагмента определяется пропорцией Маркуса-Левича-Джортнера [29]: $\ln(k_{\mathrm{ET}}) \sim \beta \cdot \Delta r$, где $\Delta r$ – длина фрагмента. Фактор $\beta$, по данным разных экспериментов, варьирует от 0,2 [30] до 1,5 [31].

В случае переноса дырки *dG* являются промежуточными акцепторами. Поэтому миграция в гетерогенной ДНК сводится к перемещениям дырки между *dG* [32, 33]. Однако, начиная с расстояния между двумя *dG* в 14 Å, пропорция Маркуса-Левича-Джортнера перестаёт соблюдаться, и скорость переноса начинает снижаться с ростом $\Delta r$ значительно медленнее [29]. Перенос при $\Delta r$ > 14 Å определяется другим механизмом – последовательными прыжками между промежуточными основаниями, разделяющими *dG* [29]. Таким образом, для гетерогенной ДНК показано как минимум два различных механизма, которые были впервые описаны Дж. Джортнером с соавт. [34]. Согласно первому – механизму суперобмена – заряд туннелируется от донора к акцептору напрямую, не взаимодействуя химически с промежуточными основаниями. Вторая, прыжковая модель, предполагает последовательный переход дырки с одного основания на другое за счёт энергии тепловых флуктуаций.

Позднее было установлено, что скорость переноса повышается с ростом температуры, а её зависимость от $\Delta r$ является периодичной [35, 36]. Авторы объясняли эти особенности возникновением специфических конформаций, способных к эффективному транспорту заряда [35]. Однако не исключено, что подобные эффекты отчасти связаны и с повышением скорости переноса по прыжковому механизму; периодичность же может быть связана со спиральной структурой ДНК.

Наиболее точные значения $k_{ET}$ отдельных стадий переноса получают в спектроскопических измерениях с разрешением по времени (англ. time-resolved methods). Самыми известными экспериментами в этой области являются работы Фредерика Льюиса с соавт. [37–41]. Благодаря небольшой длине и специфическому строению олигомеров ДНК, в них удалось измерить $k_{ET}$ миграции между парами ближайших *dG*, причём в обоих направлениях. Схема всех экспериментов была одинакова. Один из концов каждого олигомера был ковалентно связан стильбендикарбоксамидом (далее *St*). Последовательность большинства ДНК имела вид $St{-}X_n{-}dG{-}Y_m{-}dG{-}dG{-}dA_k$ или $St{-}X_n{-}dG{-}Y_m{-}dG{-}dG{-}dG{-}dA_k$, где X и Y – *dA* или *dT*, $2 \le n \le 3$; $1 \le m \le 2$; $0 \le k \le 2$. В дальнейшем мы будем нумеровать основания того участка, для которого исследуем или приводим кинетические характеристики переноса.

Фотовозбуждение *St* приводит к смещению на него электрона с ближайшего *dG* и образованию первичной ион-радикальной пары (далее ИРП) $\{St^{\bullet-}\cdot\cdot dG^{\bullet+}\}$. Таким образом, в последующем обратимом переносе заряда в направлении 5'→3' как донором, так и акцептором являются *dG*. Кинетика переноса исследуется методом спектроскопии кратковременного поглощения.

Ранее нам удалось воспроизвести экспериментальные данные Ф. Льюиса по переносу заряда в конструкте $St{-}dA{-}dA{-}dG_1{-}dA_2{-}dG_3{-}dG_4$ [37] путём вычислительного эксперимента, где исследовалась мезоскопическая модель Пейярда-Бишопа-Холстейна [42]. Однако подобные вычисления сопряжены с большими затратами машинного времени. Кроме того, они дают мало информации о физической природе переноса: более полезными в этом случае являются аналитические подходы.

Для адекватного описания транспорта заряда в $\pi$-сопряжённых системах необходимо, наряду с введением средних вероятностных характеристик системы, сохранить часть обобщенных динамических переменных. Подобные обобщенные координаты определяются гамильтонианом системы. Они меняются в зависимости от грубости усреднения и набора динамических свойств системы, представляющих интерес в каждом конкретном случае.

Пусть начальные условия являются гиббсовскими. Рассмотрим для простоты фрагмент ДНК с особой последовательностью нуклеотидов, позволяющей транспорт заряда только по одной из цепей. Преимущественное движение заряда по одной и той же цепи доказано экспериментально [43]. ДНК, в которых «перескоки» заряда между цепями крайне маловероятны, исследовались в целом ряде экспериментов [8, 44, 45]. Поэтому данное приближение является достаточно хорошим. Для изучения механизма переноса заряда в таком фрагменте можно, в данном случае, пользоваться линейным гамильтонианом, аналогичным гамильтониану Фрёлиха.

Во фрагментах ДНК электронно-колебательное взаимодействие обычно велико. Это приводит к участию в процессах переноса многих колебаний и к сильному влиянию температуры на проводимость. В нашем случае «кристалл» конечен и неоднороден. В 1959 году Холстейн рассмотрел эту задачу для одномерного однородного бесконечного кристалла и показал, что имеют место когерентный (зонный) и некогерентный (прыжковый) механизмы переноса [46]. Ниже мы увидим, что они идут параллельно и накладываются друг на друга.

Целью работы является аналитическое исследование вкладов различных механизмов переноса в транспорт заряда на небольшие расстояния в ДНК, а также определение зависимости соотношения этих вкладов от температуры. Параметры модели определялись нами из сравнения с экспериментальными данными Льюиса с соавт. [37–41].

Основой аналитических расчётов была формула Кубо для переноса заряда [47]. Однако при расчёте корреляционных функций (далее КФ), входящих в эту формулу, мы не пользовались теорией возмущения по параметру электронно-колебательного взаимодействия и по параметру интеграла переноса. Для случая ДНК этот подход оправдан, так как оба этих параметра очень велики, хотя интеграл переноса существенно уменьшается за счет поляронного эффекта.

Во второй части проведен расчет КФ токов, которые определяют тензор проводимости. Путём разложения по степеням корреляций получена замкнутая система уравнений для КФ в модели с гамильтонианом типа Фрёлиха для неоднородных фрагментов ДНК. В уравнениях были учтены только парные корреляции, что значительно упростило задачу. Проводимость исследована в модели полярона малого радиуса для неоднородного квазиодномерного фрагмента ДНК. Рассчитаны локальные константы скорости переноса заряда между сайтами и их зависимость от температуры и интенсивности взаимодействия носителей заряда с колебаниями.

В третьей части на основе сравнения результатов расчёта с экспериментальными данных Льюиса с соавт. [37–41] определены параметры модели. Изучены вклады туннельного и прыжкового механизмов, а также влияние обратного переноса заряда во фрагментах ДНК различной нуклеотидной последовательности.

## 2. ОПРЕДЕЛЕНИЕ КОЭФФИЦИЕНТА ПРОВОДИМОСТИ ПО КУБО.

Возьмём в качестве объекта моделирования B-ДНК, как наиболее часто встречающуюся форму этой молекулы. Воспользуемся формализмом матрицы плотности и приведем точное выражение для коэффициента электропроводности по формуле Кубо [47]. Тензор электропроводности в направлении $x$ равен

$$\sigma_{xx} = \frac{1}{L}\lim_{s\to 0}\int_0^{\beta} d\lambda \int_0^{\infty} e^{-st} < J_x(-i\lambda) J_x(t) > dt,$$

где скобки $<\ldots>$ означают усреднение с матрицей плотности $\rho$, $\rho = \rho' \exp[-\beta(H-F)] / Sp(\exp[-\beta(H-F)]\rho')$,

где $\rho' = T\exp\left[\sum \int_0^{\infty} dt(\eta_q^+(t) b_q(t) + \eta_q^-(t) b_q^+(t))\right]$; $T$ – оператор временного упорядочения, $L$ – длина фрагмента ДНК, $\beta$ – обратная величина температуры, $H$ – гамильтониан, $F$ – свободная энергия системы, $J(t)$ – оператор тока в гейзенберговском представлении.

Вспомогательные классические поля $\eta_q^{\pm}$ позволяют записать бесконечную цепочку дифференциальных уравнений в виде одного уравнения в функциональных производных.

Запишем гамильтониан для B-формы ДНК:

$$H=\sum_{m}\varepsilon_m a_m^+ a_m+\sum_{m}B_{m-1,m}a_{m-1}^+ a_m+ \sum_{m}B_{m+1,m}a_{m+1}^+ a_m+\sum_{q}\omega_q b_q^+ b_q-\sum_{m,q}A_{mq}\omega_q(b_q+b_{-q}^+)a_m^+ a_m, \quad (1)$$

где $m$ – номер узла, соответствующего нуклеотиду в реальной ДНК, $a_m^+(a_m)$ – операторы рождения (уничтожения) частиц электронной подсистемы, $b_q^+(b_q)$ – операторы рождения (уничтожения) частиц колебательной подсистемы, $A_{mq}$ – параметр электронно-колебательного взаимодействия. В квазиодномерном случае $A_{mq}=A_q e^{iqm}$ – безразмерные величины, характеризующие поляризацию среды в окрестности $m$-го узла; $\varepsilon_m$ – энергия электрона (дырки) на $m$-м узле, $B_{m\pm 1,;m}$ – резонансный интеграл перекрывания состояний узлов $m\pm 1$ и $m$, а $\omega_q$ – частоты колебаний.

Оператор тока равен

$$J_x=-ie\sum(x_m-x_{m+g})B_{m,m+g}a_m^+ a_{m+g} \quad (2)$$

где $x_m$ – координата узла.

Отсюда можно получить выражение для коэффициента проводимости ($L = N\cdot a$, где $N$ – число узлов фрагмента ДНК, $a$ – расстояния между сайтами)

$$\sigma_{xx}=-e^2\,\mathrm{Re}\frac{1}{L}\lim_{s\to 0}\int_0^{\beta}d\lambda\int_0^{\infty}e^{-st}\sum(x_m-x_{m+g_1})(x_n- -x_{n+g_2})<B_{m,m+g_1}a_m^+(-i\lambda)a_{m+g_1}(-i\lambda)B_{n,n+g_2}a_n^+(t)a_{n+g_2}(t)> \quad (3)$$

Определим корреляционные функции

$$G_{n,n+g}(-i\lambda;t)=<a_{m_1}^+(-i\lambda)a_{n_1}(-i\lambda)a_n^+(t)a_{n+g}(t)>=G_{n,n+g}(0;t+i\lambda),\qquad g=\pm 1,$$

Наша задача – получить уравнения для расчета КФ, входящих в (3), для выбранного гамильтониана (1) в приближении разложения по числу коррелирующих частиц. Уравнение для КФ имеет вид

$$i\frac{d}{dt}G_{m,n}(-i\lambda;t)=(-\varepsilon_m+\varepsilon_n)G_{m,n}(-i\lambda;t)-B_{m,m-1}G_{m-1,n}(-i\lambda;t)-B_{m,m+1}G_{m+1,n}(-i\lambda;t)+ +B_{n,n-1}G_{m,n-1}(-i\lambda;t)+B_{n,n+1}G_{m,n+1}(-i\lambda;t)+\sum_q\omega_q(A_{mq}-A_{nq})<(b_q+b_{-q}^+)\hat{G}_{m,n}(-i\lambda;t)> \quad (4)$$

где введено обозначение $\hat{G}_{n,n+g}(-i\lambda;t)=a_{m_1}^+(-i\lambda)a_{n_1}(-i\lambda)a_n^+(t)a_{n+g}(t)$.

Заметим, что в (4) вошли КФ более высокого порядка $<b_q\hat{G}_{m,n}(-i\lambda;t)>$ и $<b_{-q}^+\hat{G}_{m,n}(-i\lambda;t)>$. Для них следует записать отдельные уравнения, в которые войдут КФ еще более высокого порядка (и так до бесконечности). Имеем

$$i\frac{d}{dt}<b_{q_1}\hat{G}_{m,n}(-i\lambda;t)>=\omega_{q_1}<b_{q_1}\hat{G}_{m,n}>-\omega_{q_1}(A_{mq_1}-A_{nq_1})G_{m,n}(-i\lambda;t)+ (-\varepsilon_m+\varepsilon_n)<b_{q_1}\hat{G}_{m,n}(-i\lambda;t)>-B_{m,m-1}<b_{q_1}\hat{G}_{m-1,n}(-i\lambda;t)>-B_{m,m+1}<b_{q_1}\hat{G}_{m+1,n}(-i\lambda;t)>+ B_{n,n-1}<b_{q_1}\hat{G}_{m,n-1}(-i\lambda;t)>+B_{n,n+1}<b_{q_1}\hat{G}_{m,n+1}(-i\lambda;t)>+ \sum_q\omega_q(A_{mq}-A_{nq})<Tb_{q_1}(b_q+b_{-q}^+)\hat{G}_{m,n}(-i\lambda;t)> \quad (5)$$

$$i\frac{d}{dt}<b^+_{q_1}\hat{G}_{m,n}(-i\lambda;t)>=-\omega_{q_1}<b^+_{q_1}\hat{G}_{m,n}>+\omega_{q_1}(A_{mq_1}-A_{nq_1})G_{m,n}+$$
$$(-\varepsilon_m+\varepsilon_n)<b^+_{q_1}\hat{G}_{m,n}(-i\lambda;t)>-B_{m,m-1}<b^+_{q_1}\hat{G}_{m-1,n}(-i\lambda;t)>-B_{m,m+1}<b^+_{q_1}\hat{G}_{m+1,n}(-i\lambda;t)>+$$
$$B_{n,n-1}<b^+_{q_1}\hat{G}_{m,n-1}(-i\lambda;t)>+B_{n,n+1}<b^+_{q_1}\hat{G}_{m,n+1}(-i\lambda;t)>+$$
$$\sum_q\omega_q(A_{mq}-A_{nq})<Tb^+_{q_1}(b_q+b^+_{-q})\hat{G}_{m,n}(-i\lambda;t)> \quad (6)$$

В (5) и (6) возникли новые КФ, $<Tb^+_{q_1}b_{q_2}\hat{G}_{m,n}(-i\lambda;t)>$, $<Tb_{q_1}b_{q_2}\hat{G}_{m,n}(-i\lambda;t)>$, $<Tb^+_{q_1}b^+_{q_1}\hat{G}_{m,n}(-i\lambda;t)>$ и $<Tb_{q_1}b^+_{q_1}\hat{G}_{m,n}(-i\lambda;t)>$. Запишем для одной из них уравнения, которые будут иметь вид

$$i\frac{d}{dt}<Tb^+_{q_1}b_{q_2}\hat{G}_{m,n}(-i\lambda;t)>=(-\omega_{q_1}+\omega_{q_2})<Tb^+_{q_1}b_{q_2}\hat{G}_{m,n}>+$$
$$\sum_{m_1}\left(\omega_{q_1}A_{m_1q_1}<b_{q_2}a^+_{m_1}a_{m_1}\hat{G}_{m,n}>-\omega_{q_2}A_{m_1q_2}<b^+_{q_1}a^+_{m_1}a_{m_1}\hat{G}_{m,n}>\right)+$$
$$(-\varepsilon_m+\varepsilon_n)<Tb^+_{q_1}b_{q_2}\hat{G}_{m,n}(-i\lambda;t)>-B_{m,m-1}<Tb^+_{q_1}b_{q_2}\hat{G}_{m-1,n}(-i\lambda;t)>-$$
$$-B_{m,m+1}<Tb^+_{q_1}b_{q_2}\hat{G}_{m+1,n}(-i\lambda;t)>+$$
$$B_{n,n-1}<Tb^+_{q_1}b_{q_2}\hat{G}_{m,n-1}(-i\lambda;t)>+B_{n,n+1}<Tb^+_{q_1}b_{q_2}\hat{G}_{m,n+1}(-i\lambda;t)>+$$
$$\sum_q\omega_q(A_{mq}-A_{nq})<Tb^+_{q_1}b_{q_2}(b_q+b^+_{-q})\hat{G}_{m,n}(-i\lambda;t)> \quad (7)$$

Эту цепочку можно замкнуть по колебательным степеням свободы, воспользовавшись функциональными производными. Без ограничения общности вполне допустимо положить

$$<b_{q_1}(t)\hat{G}_{m,n}(t)>=\left(<b_{q_1}(t)>+\frac{\delta}{\delta\eta^+_{q_1}(t)}\right)G_{m,n}(t)=G_{m,n}(t)M_{q_1}(m_1,n_1,m,n;t) \quad (8)$$

Последнее равенство является определением функционала $M_q$. Аналогично имеем

$$<b^+_{q_1}(t)\hat{G}_{m,n}(t)>=G_{m,n}(t)M^+_{q_1}(m_1,n_1,m,n;t) \quad (9)$$

а также

$$<Tb^+_{q_1}(t)b_{q_2}(t-0)\hat{G}_{m,n}(t)>=G_{m,n}(t)(M^+_{q_1}(m_1,n_1,m,n;t)+\frac{\delta}{\delta\eta^-_{q_1}(t)})M^+_{q_2}(m_1,n_1,m,n;t) \quad (10)$$

$$<Tb^+_{q_1}(t)b_{q_2}(t-0)b^x_{q_3}(t-0-0)\hat{G}_{m,n}(t)>=G_{m,n}(t)(M^+_{q_1}(m_1,n_1,m,n;t)+$$
$$\frac{\delta}{\delta\eta^-_{q_1}(t)})(M^+_{q_2}(m_1,n_1,m,n;t)+\frac{\delta}{\delta\eta^+_{q_2}(t)}) \quad (11)$$

Выражения (10) и (11) позволяют свернуть бесконечную цепочку уравнений КФ и получить методы ее аппроксимации. Действительно, действуя оператором вариационной производной на функционалы и переходя к пределу $\eta^x_{qj}(t)\to 0$, из уравнения (9) можно получить выражения, не содержащие классических полей

$$<b^+_{q_1}(t)\hat{G}_{m,n}(t)>=M^+_{q_1}(m_1,n_1,m,n;t)G_{m,n}(t) \quad (12)$$

Теперь $G_{m,n}(t)$, $M_q^+(m_1,n_1,m,n;t)$ и $M_q(m_1,n_1,m,n;t)$ превращаются из функционалов в обычные функции. Выражение (12) означает, что действие оператора $b_{q_1}^+(t)$ внутри усреднения заменяется эквивалентным по действию внешним полем, зависящим от времени, выбора системы и от типа рассматриваемой КФ. Для оператора уничтожения имеем, соответственно

$$< b_{q_1}(t)\hat{G}_{m,n}(t) >= M_{q_1}(m_1,n_1,m,n;t)G_{m,n}(t),$$

Поле $M_{q_1}^+(m_1,n_1,m,n;t)$ отличается от поля $M_{q_1}(m_1,n_1,m,n;t)$, но является пока неопределенным, как и поле $M_{q_1j_1}(m_1,n_1,m,n;t)$. Далее мы получим соответствующие уравнения для их определения. При появлении двух колебательных состояний (операторов) возникает корреляция между двумя полями. Она учтена через второе слагаемое правой части в выражении (10). Переходя в этом уравнении к нулевому пределу по классическим полям, имеем

$$< Tb_{q_1}^+(t)b_{q_2}(t-0)\hat{G}_{m,n}(t) >= \Big(M_{q_1}^+(m_1,n_1,m,n;t)M_{q_1}(m_1,n_1,m,n;t)+$$
$$+\frac{\delta}{\delta\eta_{q_1}^-(t)}M_{q_2}(m_1,n_1,m,n;t)\Big)G_{m,n}(t), \qquad \{\eta_q^x\}\to 0 \quad .$$

$M_{q_1}^+(m_1,n_1,m,n;t)$ и $M_{q_1}(m_1,n_1,m,n;t)$ по физическому смыслу являются полями, действующими на частицу (носитель тока) без учета корреляций между ними. Следовательно, учет парных корреляций между этими полями отражает парные корреляции в колебательной подсистеме. Обозначим эти парные корреляции как

$$D_{q_1;q_2}^{+-}(m_1,n_1,m,n;t)=\frac{\delta}{\delta\eta_{q_1}^-(t)}M_{q_2}(m_1,n_1,m,n;t), \qquad \{\eta_q^x\}\to 0,$$

Если учитывать только парные корреляции между колебаниями за счет нелинейных взаимодействий или взаимодействий через промежуточную частицу носителя, то получим

$$< Tb_{q_1}^+(t)b_{q_2}(t-0)\hat{G}_{m,n}(t) >= (M_{q_1}^+(t)M_{q_1}(t)+D_{q_1,q_2}^{+-}(t))G_{m,n}(t),$$

Здесь и далее, в целях упрощения записи, принято тождественно $M_{q_1}^+(m_1,n_1,m,n;t)$ $=M_{q_1}^+(t)$, $M_{q_1}(m_1,n_1,m,n;t)=M_{q_1}(t)$ и $D_{q_1;q_2}^{+-}(m_1,n_1,m,n;t)=D_{q_1;q_2}^{+-}(t)$.

Величины $M_{q_1}^+(t)$ и $M_{q_1}(t)$ учитывают влияние на носитель от первой моды, а $M_{q_2}^+(t)$ и $M_{q_2}(t)$ – от второй моды. Возможность согласованных движений первой и второй мод не рассматривается; $D_{q_1;q_2}^{+,-}(t)$ учитывает возможность корреляций между парами мод при взаимодействии с носителями. В этом приближении учтены все парные корреляции в колебательной подсистеме.

В случае с тремя модами, когда возможны парные корреляции первой моды со второй, второй – с третьей и первой с третьей, а также тройные корреляции, имеем

$$< Tb_{q_1}^+(t)b_{q_2}(t)b_{q_3}^x(t)\hat{G}_{m,n}(t) >= \Big(M_{q_1}^+(t)M_{q_2}(t)M_{q_3}^x(t)+M_{q_1}^+(t)D_{q_2;q_3}^{-,x}(t)+$$
$$+M_{q_2}(t)D_{q_1;q_3}^{+,x}(t)+M_{q_3}^x(t)D_{q_1;q_2}^{+,-}(t)+\frac{\delta}{\delta\eta_{q_1}^-(t)}D_{q_2;q_3}^{-,x}(t)\Big)G_{m,n}(t), \qquad \{\eta_q^x\}\to 0'$$

Выражение

$$\frac{\delta}{\delta\eta^{-}_{q_1}(t)} D^{\pm,\pm}_{q_2;q_3}(t) = T^{+,\pm,\pm}_{q_1;q_2;q_3}$$

описывает влияние тройных корреляций на перемещение носителей тока.

Подобные корреляции несущественны, поскольку одновременное столкновение трёх частиц в рассмотренном нами квазиодномерном случае почти невозможно. Поэтому учёт корреляций выше второго порядка лишён смысла. Исключение корреляции более высоких порядков делает цепочку систем уравнений конечной. В то же время, наша методика может быть применена и к трёхмерному случаю, поскольку позволяет рассматривать корреляции любого порядка.

Численное исследование полученных систем нелинейных дифференциальных уравнений позволяет оценить погрешность как функцию максимального учтённого порядка корреляции. Изложенный метод получения замкнутых систем уравнений применим и для нелинейных взаимодействий, например для гамильтониана Пейярда-Бишопа [48].

Полагая $T^{\pm,\pm,\pm}_{q_1;q_2;q_3} = 0$, произведем обрыв цепочки уравнений. С учетом принятых обозначений и приближений перепишем уравнения (4) в виде

$$\begin{aligned} i\frac{d}{dt}G_{m,n}(t) &= (-\varepsilon_m + \varepsilon_n)G_{m,n}(t) - B_{m,m-1}G_{m-1,n}(t) - B_{m,m+1}G_{m+1,n}(t) + \\ &+ B_{n,n-1}G_{m,n-1}(t) + B_{n,n+1}G_{m,n+1}(t) + \sum_q \omega_q(A_{mq} - A_{nq})(M_q + M^{+}_{-q})G_{m,n}(t) \end{aligned} \tag{13}$$

Исключив выражение (13) из уравнения (5), получим для $M_q$ окончательно:

$$\begin{aligned} i\frac{d}{dt}M_{q_1} &= \omega_{q_1}M_{q_1} - \omega_{q_1}(A_{mq_1} - A_{nq_1}) + \sum_q \omega_q(A_{mq} - A_{nq})(D^{-,-}_{q_1;q} + D^{-,+}_{q_1;q}) + \\ &+ \{-B_{m,m-1}(M_{q_1}(m_1,n_1,m-1,n;t) - M_{q_1})G_{m-1,n} - \\ &- B_{m,m+1}(M_{q_1}(m_1,n_1,m+1,n;t) - M_{q_1})G_{m+1,n} + \\ &+ B_{n,n-1}(M_{q_1}(m_1,n_1,m,n-1;t) - M_{q_1})G_{m,n-1} + \\ &+ B_{n,n+1}(M_{q_1}(m_1,n_1,m,n+1;t) - M_{q_1})G_{m,n+1}\} \cdot G^{-1}_{m,n}(t) \end{aligned} \tag{14}$$

Аналогичные операции проведём для уравнений (6) и (7), которые определяют функции $M^{+}_q(t)$ и $D^{\pm\pm}_{q_1q_2}(t)$. Ввиду громоздкости выражений мы их не приводим: они будут даны ниже, после упрощения. Члены в фигурных скобках можно учесть итеративно, для получения решений нужной точности.

Проведем дальнейшие упрощения системы уравнений. Выражения в фигурных скобках, аналогичных скобкам в уравнении (14), учитывают изменение частот и смещение минимумов термов от присутствия в системе носителей тока. Носителей тока в системе мало – всего по одному на каждый фрагмент ДНК. Поэтому их влияние на большое число колебаний приводит к незначительным поправкам, пропорциональным $N^{-1}$, где $N$ – количество колебательных мод. Однако и эти поправки могут быть учтены методом последовательных приближений. Это позволит воспроизвести тонкие эффекты строения ДНК и выявить области, где эти поправки окажутся существенными.

В первом приближении все фигурные скобки, аналогичные скобкам в выражении (14), могут быть опущены. В этом случае уравнения для функций $G_{m,n}(t)$, $M^{+}_q(t)$, $M_q(t)$ и $D^{\pm,\pm}_{q_1;q_2}(t)$ существенно упрощаются, принимая вид

$$i\frac{d}{dt}G_{m,n}(t)=(-\varepsilon_m+\varepsilon_n)G_{m,n}(t)-B_{m,m-1}G_{m-1,n}(t)-B_{m,m+1}G_{m+1,n}(t)+ \\ +B_{n,n-1}G_{m,n-1}(t)+B_{n,n+1}G_{m,n+1}(t)+\sum_q \omega_q(A_{mq}-A_{nq})(M_q+M^+_{-q})G_{m,n}(t) \tag{15}$$

$$i\frac{d}{dt}M_{q_1}=\omega_{q_1}M_{q_1}-\omega_{q_1}(A_{mq_1}-A_{nq_1})+\sum_q\omega_q(A_{mq}-A_{nq})(D^{-,-}_{q_1;q}+D^{-,+}_{q_1;q}) \tag{16}$$

$$i\frac{d}{dt}M^+_{q_1}=-\omega_{q_1}M^+_{q_1}+\omega_{q_1}(A_{mq_1}-A_{nq_1})+\sum_q\omega_q(A_{mq}-A_{nq})(D^{+,-}_{q_1;q}+D^{+,+}_{q_1;q}) \tag{17}$$

$$i\frac{d}{dt}D^{\mp,\pm}_{q_1;q_2}(t)=(\pm\omega_{q_1}\mp\omega_{q_2})D^{\mp,\pm}_{q_1;q_2}(t) \tag{18}$$

Зададим начальные условия: $D^{+,-}_{q_1;q_2}(0)=\delta_{q_1q_2}N_{q_1}$, $D^{-,+}_{q_1;q_2}(0)=\delta_{q_1q_2}(N_{q_1}+1)$, $D^{+,+}_{q_1;q_2}(0)=0$, $D^{-,-}_{q_1;q_2}(0)=0$, где $N_q=<b^+_qb_q>$ – равновесная заселенность колебательных состояний. С учётом этих условий, уравнения (18) имеют решения

$$\begin{aligned}&D^{+,-}_{q_1;q_2}(t)=\delta_{q_1q_2}N_{q_1}\\ &D^{-,+}_{q_1;q_2}(t)=\delta_{q_1q_2}(N_{q_1}+1)\\ &D^{+,+}_{q_1;q_2}(t)=0, \qquad D^{-,-}_{q_1;q_2}(t)=0\end{aligned} \tag{19}$$

Подставляя полученные решения в (16) и (17), имеем

$$M_q(t)=e^{-i\omega_qt}(A_{mq}-A_{nq})(N_q+1)-(A_{mq}-A_{nq})(N_q+1) \tag{20}$$

$$M^+_q(t)=e^{i\omega_qt}\{-(A_{mq}-A_{nq})N_q\}+(A_{mq}-A_{nq})N_q \tag{21}$$

После того как величины $M_q(t)$ и $M^+_q(t)$ получают аналитическое выражение, исследование системы уравнений КФ сводится к численному решению линейного уравнения (15) с переменными коэффициентами.

Для случая поляронов малого радиуса имеем

$$i\frac{d}{dt}M_{q_1}=\omega_{q_1}M_{q_1}-\omega_{q_1}A_{nq_1}+\omega_{q_1}(A_{mq_1}-A_{nq_1})(N_{q_1}+1), \qquad M_q(0)=A_{nq},$$

$$i\frac{d}{dt}M^+_{q_1}=-\omega_{q_1}M^+_{q_1}+\omega_{q_1}A_{nq_1}+\omega_{q_1}(A_{mq_1}-A_{nq_1})N_{q_1}, \qquad M^+_q(0)=A_{nq}$$

Решением этих уравнений будет

$$M_q(t)=e^{-i\omega_qt}\left\{M_q(0)-i\omega_q(A_{mq}-A_{nq})\int_0^t e^{i\omega_qt_1}(N_q+1)dt_1+i\omega_q\int_0^t e^{i\omega_qt_1}A_{nq}dt_1\right\}= \\ =e^{-i\omega_qt}\left\{(A_{mq}-A_{nq})(N_q+1)\right\}-(A_{mq}-A_{nq})(N_q+1)+A_{nq} \tag{22}$$

Аналогично получим

$$M^+_q(t)=e^{i\omega_qt}\left\{-(A_{mq}-A_{nq})N_q\right\}+A_{nq}+(A_{mq}-A_{nq})N_q,$$

Подставляя полученное решение (22) в (15), получаем для линейной одномерной цепочки из *N* узлов (сайтов) систему уравнений, аналогичных уравнениям Матье

$$i\frac{d}{dt}G_{m,m}(t) = -B_{m,m-1}G_{m-1,m}(t) - B_{m,m+1}G_{m+1,m}(t) + B_{m,m-1}G_{m,m-1}(t) + B_{m,m+1}G_{m,m+1}(t)$$

$$\begin{aligned}
&i\frac{d}{dt}G_{m-1,m}(t) = (-\varepsilon_{m-1} + \varepsilon_m)G_{m-1,m}(t) - B_{m-1,m}G_{m,m}(t) + B_{m,m-1}G_{m-1,m-1}(t) + \\
&+\sum_q \omega_q \big(e^{-i\omega_q t}\{(A_{m-1q} - A_{mq})^2(N_q+1)\} - (A_{m-1q} - A_{mq})^2 + 2A_{mq}(A_{m-1q} - A_{mq}) \\
&+e^{i\omega_q t}\{-(A_{m-1q} - A_{mq})^2 N_q\}\big)G_{m-1,m}(t) \qquad (23)\\
&i\frac{d}{dt}G_{m+1,m}(t) = (-\varepsilon_{m+1} + \varepsilon_m)G_{m+1,m}(t) - B_{m+1,m}G_{m,m}(t) + B_{m,m}G_{m+1,m+1}(t) + \\
&+\sum_q \omega_q \big(e^{-i\omega_q t}\{(A_{m+1q} - A_{mq})^2(N_q+1)\} - (A_{m+1q} - A_{mq})^2 + 2A_{mq}(A_{m+1q} - A_{mq}) \\
&+e^{i\omega_q t}\{-(A_{m+1q} - A_{mq})^2 N_q\}\big)G_{m+1,m}(t), \qquad m = 1,2,\dots,N
\end{aligned}$$

с начальными условиями $G_{m,n}(0) = \delta_{n_1 m}\delta_{m_1 n}n_n$, где $\delta_{n_1 m}$ – символ Кронекера, а $n_n$ – заселённость $n$-го узла носителями тока в начальный момент времени.

Выражение (23) включает две различных категории переходов. Первая категория – бесфононные процессы, которые имеют максимум для однородного случая при нулевой температуре, когда $\sum N_q = \sum N_{q_1} = 0$. Вторая – фононные процессы, когда $\sum N_q \neq \sum N_{q_1}$.

Рассмотрим систему (23) для случая четырех сайтов ($m$ = 0, 1, 2, 3). Проведем расчеты токовой КФ для перехода дырки с сайта 1 на сайт 2, то есть вычислим КФ $G_{21}(t)$. Согласно формуле Кубо, для вероятности локального перехода $\sigma_{1\to 2}$, в приближении единственной колебательной частоты, получим

$$\begin{aligned}
\sigma_{1\to 2} &= -e^2\,\mathrm{Re}\frac{1}{L}\lim_{s\to 0}\int_0^{\beta} d\lambda \int_0^{\infty} e^{-st}(x_1 - x_2)^2 B_{1,2}^2 < a_1^+(-i\lambda)a_2(-i\lambda)a_2^+(t)a_1(t) > dt = \\
&= -e^2\,\mathrm{Re}\frac{1}{L}\lim_{s\to 0}\int_0^{\beta} d\lambda \int_0^{\infty} e^{-st}(x_1 - x_2)^2 B_{1,2}^2 G_{21}(t + i\lambda)dt
\end{aligned} \qquad (24)$$

Величина $\sigma_{1\to 2}(T)$ была рассчитана численно, по формуле (24), для ряда значений параметров соседних с ними сайтов 0 и 3, которые приведены в Таблице 1. При этом значения величин для сайтов 1 и 2 составляли: $\omega_1 = \omega_2$ = 0,3; $B_{1,2} = B_{2,1}$ = 0,5; $(A_1 - A_2)^2$ = 16; $\sigma_0 = e^2 n_2 B_{12}^2 (x_1 - x_2)^2$; $E_{21} = -E_{12} = -\varepsilon_1 + \omega_1 A_1^2 + \varepsilon_2 - \omega_2 A_2^2$ = 3,0.

**Таблица 1.** Набор параметров для сайтов 0 и 3.

| № | $\omega_0$ | $\omega_3$ | $E_{01}$ | $E_{23}$ | $(A_0 - A_1)^2$ | $(A_2 - A_3)^2$ |
|---|---|---|---|---|---|---|
| 1 | 0,3 | 0,7 | –1 | –3 | 4 | 2 |
| 2 | 0,4 | 0,7 | –1 | 3 | 4 | 6 |
| 3 | 0,5 | 0,6 | 1 | 9 | 3 | 2 |
| 4 | 0,6 | 0,5 | 2 | 3 | 2 | 16 |

Как видно из Таблицы 1, параметры для сайтов 0 и 3 варьировались в широком диапазоне. Несмотря на это, максимальная разность соотношения $\sigma_{1\to 2}(T)/\sigma_0$ для серии соответствующих им функций не превышала $10^{-6}$ в интервале от 50 до 400 К.

Далее была исследована температурная зависимость переноса заряда. Основным фактором, определяющим величину $\sigma_{1\to 2}(T)/\sigma_0$, оказался квадрат разности параметров электронно-колебательного взаимодействия сайтов, между которыми происходит перенос, $(A_1 - A_2)^2$. На Рис. 1 приведены результаты расчётов для значений этой разности от 0,7 до 10 безразмерных единиц. При любых $(A_1 - A_2)^2$ функции $\sigma_{1\to 2}(T)/\sigma_0$ зависели только от свойств сайтов 1 и 2. Их изменения, вызванные варьированием параметров для сайтов 0 и 3, были исчезающе малы, см. выше.

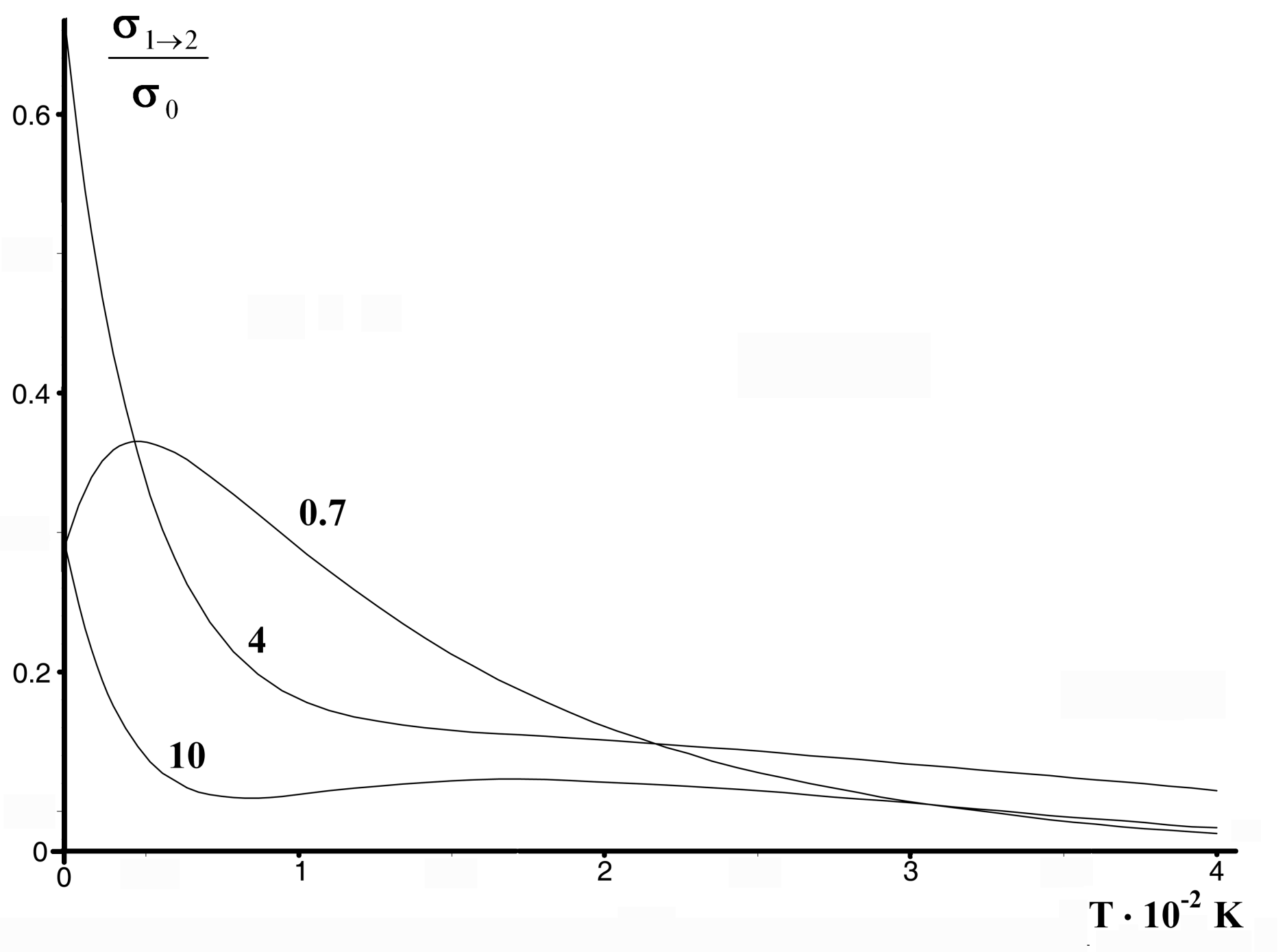

**Рис. 1**. Изменение температурной зависимости относительной локальной проводимости звена 1–2 при различных значениях $(A_1 - A_2)^2$, указанных рядом с соответствующими кривыми; $E_{01} = -1$; $E_{23} = -3$; $(A_0 - A_1)^2 = 4$; $(A_2 - A_3)^2 = 2$; $\omega_0 = 0,3$; $\omega_3 = 0,7$.

Из Рис. 1 видно, что для малых $(A_1 - A_2)^2$ механизм проводимости – зонный, с ростом температуры она убывает. При низких T также возможно снижение проводимости, связанное с эффектами конечного размера фрагмента ДНК. Оно связано с приближением единственной колебательной частоты: при конечном фрагменте не всегда удается добиться резонанса начального и конечного состояний.

С увеличением $(A_1 - A_2)^2$, при низких температурах растёт вклад проводимости по прыжковому механизму. Повышение T ведёт к дальнейшему росту этого вклада, который, начиная с некоторой температуры, становится определяющим. При повышении T выше некоторого критического значения, скорость переноса по этому механизму также начинает снижаться.

Сравним результаты, полученные при помощи нашего подхода, с проводимостью бесконечных однородных кристаллов, вычисленной Холстейном [46]. Опуская все резонансные интегралы в выражении (15), получаем решение в виде

$$G_{m,n}(t) = \delta_{m_1 n}\delta_{n_1 m} n_n \exp\left\{ i\sum_{qj} 4\omega_{qj} A_{qj}^2 (1-\cos q)t + \right. \\ \left. + \sum_{qj} 2A_{qj}^2 (1-\cos q)[N_q(e^{i\omega_{qj}t} - 1) + (N_{qj}+1)(e^{-i\omega_{qj}t} - 1)] \right\} \quad (25)$$

Выражение для проводимости, с учетом (25), имеет вид

$$\sigma_{xx} = -\mathrm{Re}\, e^2 \frac{1}{L} \lim_{s\to 0} \int_0^{\beta} d\lambda \int_0^{\infty} dt e^{-s(t+i\lambda)} \sum (x_n - x_m)^2 B_{n,m} B_{m,n} n_n \cdot$$

$$\cdot \exp\left\{ \sum_q -2A_q^2 (1-\cos q)(2N_q+1) \right\} \cdot \exp\left\{ i\sum_q 4A_q^2 \omega_q (1-\cos q)(t+i\lambda) + \right. \tag{26}$$

$$\left. + \sum_q 2A_q^2 (1-\cos q) \left[ N_q e^{i\omega_q (t+i\lambda)} + (N_q+1) e^{-i\omega_q (t+i\lambda)} \right] \right\}$$

В приближении единственной частоты колебаний выражение (26) принимает вид

$$\sigma_{xx} = -\mathrm{Re}\, e^2 \frac{1}{L} \lim_{s\to 0} \int_0^{\beta} d\lambda \int_0^{\infty} dt e^{-s(t+i\lambda)} \sum (x_n - x_m)^2 B_{n,m} B_{m,n} n_n \exp\{-\gamma(2N+1)\} \times$$

$$\times \sum_{p,k} \frac{(\gamma N)^p}{p!} \frac{\gamma^k (N+1)^k}{k!} e^{i(2\gamma\omega - \omega k + \omega p)(t+i\lambda)} \tag{27}$$

где $\gamma = 2A_{q_0}^2 (1-\cos q_0)$.

В случае однородной системы формула (27) совпадает с соответствующими выражениями Холстейна для проводимости [46]. Для выяснения физических основ переноса имеет смысл, в частности, рассчитать $\sigma_{xx}$ в предельных случаях высоких и низких температур, а также слабого и сильного электронно-колебательного взаимодействия. При низких T можно провести разложение по параметру электронно-колебательного взаимодействия, а для высоких – по времени. В последнем случае теряются все особенности, связанные с гетерогенностью системы. Низкотемпературное разложение проводится с участием одного, двух, трех и т.д. колебательных процессов, в которых участвуют разные моды и выполняются законы сохранения энергии. При этом остаются только члены с комбинациями частот, в сумме равными нулю. При высоких T эти процессы носят активационный характер, поскольку в них участвует множество колебательных мод.

## 3. РЕЛАКСАЦИЯ ДЫРКИ ВО ФРАГМЕНТАХ ДНК ВИДА $G(A)_n(G)_m$

Перейдём от описания методики и исследования поведения системы в общем виде к расчёту переноса в конкретных фрагментах ДНК. Воспользовавшись методом, изложенным в части 2, получим замкнутую систему уравнений для определения заселенностей узлов. Для двух сайтов она принимает вид

$$i\frac{d}{dt} G_{11}(t) = -B_{12} G_{21}(t) + B_{12} G_{12}(t) + i\alpha G_{11}(t), \qquad G_{11}(0) = 1 \tag{28}$$

$$i\frac{d}{dt} G_{21}(t) = (E_{12} + F_{12}(t)) G_{21}(t) + B_{12}(G_{22}(t) - B_{12} G_{11}(t)), \qquad G_{21}(0) = 0 \tag{29}$$

$$i\frac{d}{dt} G_{12}(t) = (-E_{12} + F_{12}(t)) G_{12}(t) - B_{12}(G_{22}(t) - G_{11}(t)), \qquad G_{12}(0) = 0 \tag{30}$$

$$i\frac{d}{dt} G_{22}(t) = B_{12}(G_{21}(t) - G_{12}(t)), \qquad G_{22}(0) = 0 \tag{31}$$

где $G_{mn}(t) = < a_m^+(t) a_n(t) >$ – корреляционная функция операторов дырки, $<\ldots>$ – символ квантового и статистического усреднения по начальным распределениям колебательных состояний, $N=(e^{\beta\omega}-1)^{-1}$ – число заполнения колебательного состояния, $\alpha$ – скорость рекомбинации ион-радикальной пары, $\beta=(kT)^{-1}$ – обратная величина температуры, $k$ – постоянная Больцмана в эВ/К, $T$ – абсолютная температура, $\omega$ – круговая частота колебаний в узле, $B_{12}$ – матричный элемент переноса между узлами, $F_{12}(t)=F_{21}(t)=\omega(A_1-A_2)^2[Ne^{-i\omega t}-(N+1)e^{+i\omega t}]$, а $E_{12}=-\varepsilon_1+\omega A_1^2+\varepsilon_2-\omega A_2^2=-E_{21}$. Выбранные значения параметров: $B_{12}$ = 0,089 эВ [49]; $\varepsilon_1$ = 1,24 эВ, $\varepsilon_2$ = 1,69 эВ [18]; $\omega$ = 0,001 эВ, $\alpha$ = 4 · $10^{-7}$ эВ. Последние две величины вычислены, соответственно, по характеристической частоте колебаний нуклеотидной пары вдоль водородной связи [42] и времени рекомбинации ИРП $\{St^{\bullet-}\cdot\cdot dG^{\bullet+}\}$ в олигомере $St$–$dA$–$dA$–$dG_1$–$dA_2$–$dG_3$–$dG_4$ [37–41].

Для выявления вклада разных механизмов в процесс переноса, получим сначала приближенное аналитическое решение системы (28) – (31). Поскольку $\alpha << 1$, а $\frac{dF_{12}}{dt} << F_{12}^2$, то систему можно переписать в виде одного уравнения

$$\left(\left(i\frac{d}{dt}-F_{12}(t)\right)^2-E_{12}^2\right)i\frac{d}{dt}G_{11}(t)=-2B_{12}^2\left(i\frac{d}{dt}-F_{12}(t)\right)(1-2G_{11}(t)),$$

Теперь, если пренебречь в правой части членом $F_{12}(t)$, получим аналитически решаемое уравнение:

$$\left(\left(i\frac{d}{dt}-F_{12}(t)\right)^2-E_{12}^2-4B_{12}^2\right)i\frac{d}{dt}G_{11}(t)=0,$$

с решением

$$G_{11}(t)=C_0+\exp\left(-i\int_0^t F_{12}(t_1)dt_1\right)\cdot\left(C_1\exp\left(i\sqrt{E_{12}^2+4B_{12}^2}\,t\right)+C_2\exp\left(-i\sqrt{E_{12}^2+4B_{12}^2}\,t\right)\right) \quad (32)$$

и начальными условиями

$$G_{11}(0)=1,\qquad i\frac{d}{dt}G_{11}(0)=0,\qquad (i\frac{d}{dt})^2G_{11}(0)=2B_{12}^2,$$

Замечая, что

$$\exp(-i\int_0^t F_{12}(t_1)dt_1)=\exp(-(2N+1)(A_1-A_2)^2)\exp((A_1+A_2)^2(Ne^{-i\omega t}+(N+1)e^{i\omega t})),$$

$$F_{12}(0)=-\omega(A_1-A_2)^2,\qquad i\frac{d}{dt}F_{12}(0)=-\omega^2(A_1-A_2)^2(2N+1)$$

определим постоянные коэффициенты, составляющие, соответственно

$$C_0=1-\frac{2B_{12}^2}{E_{12}^2+4B_{12}^2-\omega^2((A_1-A_2)^4+(A_1-A_2)^2(2N+1))},$$

$$C_1=\frac{B_{12}^2}{E_{12}^2+4B_{12}^2-\omega^2((A_1-A_2)^4+(A_1-A_2)^2(2N+1))}\left(1-\frac{\omega\cdot(A_1-A_2)^2}{\sqrt{E_{12}^2+4B_{12}^2}}\right),$$

$$C_2 = \frac{B_{12}^2}{E_{12}^2 + 4B_{12}^2 - \omega^2((A_1 - A_2)^4 + (A_1 - A_2)^2(2N+1))}\left(1 + \frac{\omega\cdot(A_1 - A_2)^2}{\sqrt{E_{12}^2 + 4B_{12}^2}}\right),$$

Таким образом, заселенность первого сайта изменяется во времени по закону

$$\begin{aligned} G_{11}(t) = C_0 + \exp\left(-(2N+1)(A_1 - A_2)^2\right)\cdot \\ \cdot\exp\left((A_1 + A_2)^2\left(Ne^{-i\omega t} + (N+1)e^{i\omega t}\right)\right)\cdot\left(C_1 e^{i\sqrt{E_{12}^2+4B_{12}^2}t} + C_2 e^{-i\sqrt{E_{12}^2+4B_{12}^2}t}\right)\end{aligned} \tag{33}$$

то есть в очень широком диапазоне частот. Это приводит к трудностям при численном расчете.

Чтобы вычисления позволяли оценить кинетические параметры переноса, необходимо провести усреднение по быстрым колебаниям. Последние связаны с недиагональными члены матрицы корреляционной функции $G_{mn}(t)$. Поэтому усреднять по времени необходимо именно эти члены. Выразим $G_{21}(t)$ и $G_{12}(t)$ из уравнений (29) и (30), вынося медленно меняющиеся функции $G_{nn}(t)$ из-под интеграла и устремляя пределы интегрирования к бесконечности. Тогда имеем

$$\begin{aligned} G_{21}(t) = -iB_{12}\left(G_{22}(t) - G_{11}(t)\right)\int_{-\infty}^{\infty}\exp\left(-iE_{12}t' - i\int_0^{t'}F_{12}(t'')dt''\right)dt' \\ G_{12}(t) = iB_{12}\left(G_{22}(t) - G_{11}(t)\right)\int_{-\infty}^{\infty}\exp\left(iE_{12}t' - i\int_0^{t'}F_{12}(t'')dt''\right)dt' \end{aligned}.$$

Подставляя полученные значения в (28), получим кинетическое уравнение

$$\begin{aligned} \frac{d}{dt}G_{11}(t) = 2B_{12}^2\left(G_{22}(t) - G_{11}(t)\right)\int_{-\infty}^{\infty}\exp\left(-i\int_0^{t'}F_{12}(t'')dt''\right)\cos(E_{12}t')dt' = \\ = k_{12}\left(G_{22}(t) - G_{11}(t)\right)\end{aligned}, \tag{34}$$

где $k_{12}$ – константа скорости переноса дырки с сайта 1 на сайт 2, усредненная по быстрым квантовым переходам. Константу $k_{12}$ определим не по приближенной формуле (34), а по более точной формуле (15)

$$k_{12} = B_{12}^2\,\mathrm{Re}\int_0^{\infty} < a_1^+(0)a_2(0)a_2^+(t)a_1(t) > dt,$$

Окончательно, выделив реальную часть, имеем для аналитического решения

$$\begin{aligned} k_{12} = \frac{2B_{1,2}^2}{4B_{1,2}^2 + E_{12}^2}e^{-(A_1-A_2)^2(2N+1)}I_0\left((A_1 - A_2)^2 2\sqrt{N(N+1)}\right) \\ \frac{1}{2}\frac{2B_{1,2}^2 + E_{12}^2 - E_{12}\sqrt{4B_{1,2}^2 + E_{12}^2}}{(4B_{1,2}^2 + E_{12}^2)}\left(\frac{N}{N+1}\right)^{\omega^{-1}\sqrt{4B_{1,2}^2+E_{12}^2}}\cdot \\ e^{-(A_1-A_2)^2(2N+1)}I_{\omega^{-1}\sqrt{4B_{1,2}^2+E_{12}^2}}\left((A_1 - A_2)^2 2\sqrt{N(N+1)}\right) + \\ \frac{1}{2}\frac{2B_{1,2}^2 + E_{12}^2 + E_{12}\sqrt{4B_{1,2}^2 + E_{12}^2}}{(4B_{1,2}^2 + E_{12}^2)}\left(\frac{N}{N+1}\right)^{\omega^{-1}\sqrt{4B_{1,2}^2+E_{12}^2}}\cdot \\ e^{-(A_1-A_2)^2(2N+1)}I_{\omega^{-1}\sqrt{4B_{1,2}^2+E_{12}^2}}\left((A_1 - A_2)^2 2\sqrt{N(N+1)}\right)\end{aligned} \tag{35}$$

где $I_\nu(z)$ – функция Бесселя от мнимого аргумента, $\nu = \omega^{-1} \cdot \sqrt{4B_{1,2}^2 + E_{12}^2}$ .

Численное решение системы уравнений (28) – (31) без приближений дало результаты, при Т > 50 К весьма близкие к аналитическим расчётам. Максимальное различие соотношения $\sigma_{1\to2}(T)/\sigma_0$ не превышало 12%.

На Рис. 2. приведены соответствующие графики. Значения $A_1$ и $A_2$, использованные при вычислениях, составляли, соответственно, 3 и 8. Процедура их определения будет описана несколько ниже.

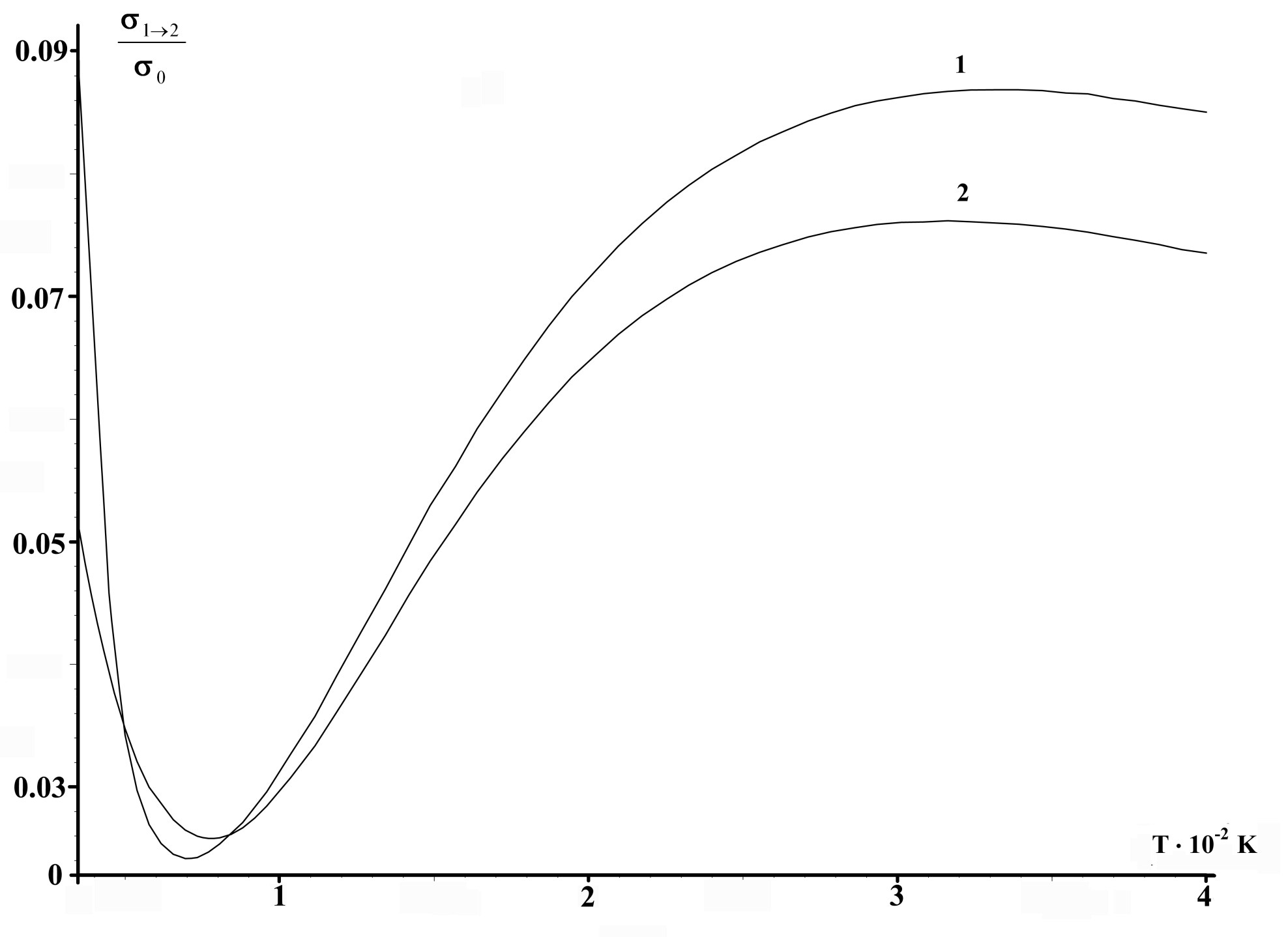

**Рис. 2**. Сравнение результатов аналитического и численного решения системы уравнений (28) – (31): 1 – кривая, полученная аналитически, 2 – численно. $B_{12}$ = 0,089 эВ; $\varepsilon_1$ = 1,24 эВ; $\varepsilon_2$ = 1,69 эВ; $\omega$ = 0,01 эВ; $A_1$ = 3; $A_2$ = 8.

На Рис. 2 также хорошо заметно качественное сходство полученных функций, в особенности близость положений их экстремумов. Следовательно, использование приближённого аналитического решения в данной задаче вполне допустимо. Более того, в отличие от численного исследования, аналитическое решение позволяет оценить вклады различных механизмов переноса по отдельности.

На Рис. 3 показан результат анализа этих вкладов в процесс переноса. Видно, что вероятность переноса по прыжковому механизму достигает своего максимума при более высоких Т, по сравнению с туннельным механизмом. При низких температурах скорость миграции заряда снижена за счёт «обратного переноса», обусловленного отражением дырки от активационного барьера «прыжка». Вероятность отражения показана кривой 3 на Рис. 3. Низкая скорость транспорта заряда при малых температурах качественно согласуется с экспериментальными данными [50]. Кроме того, преобладание переноса по прыжковому механизму при достаточно высоких температурах было показано ранее в теоретических работах других исследователей, см., напр., [51].

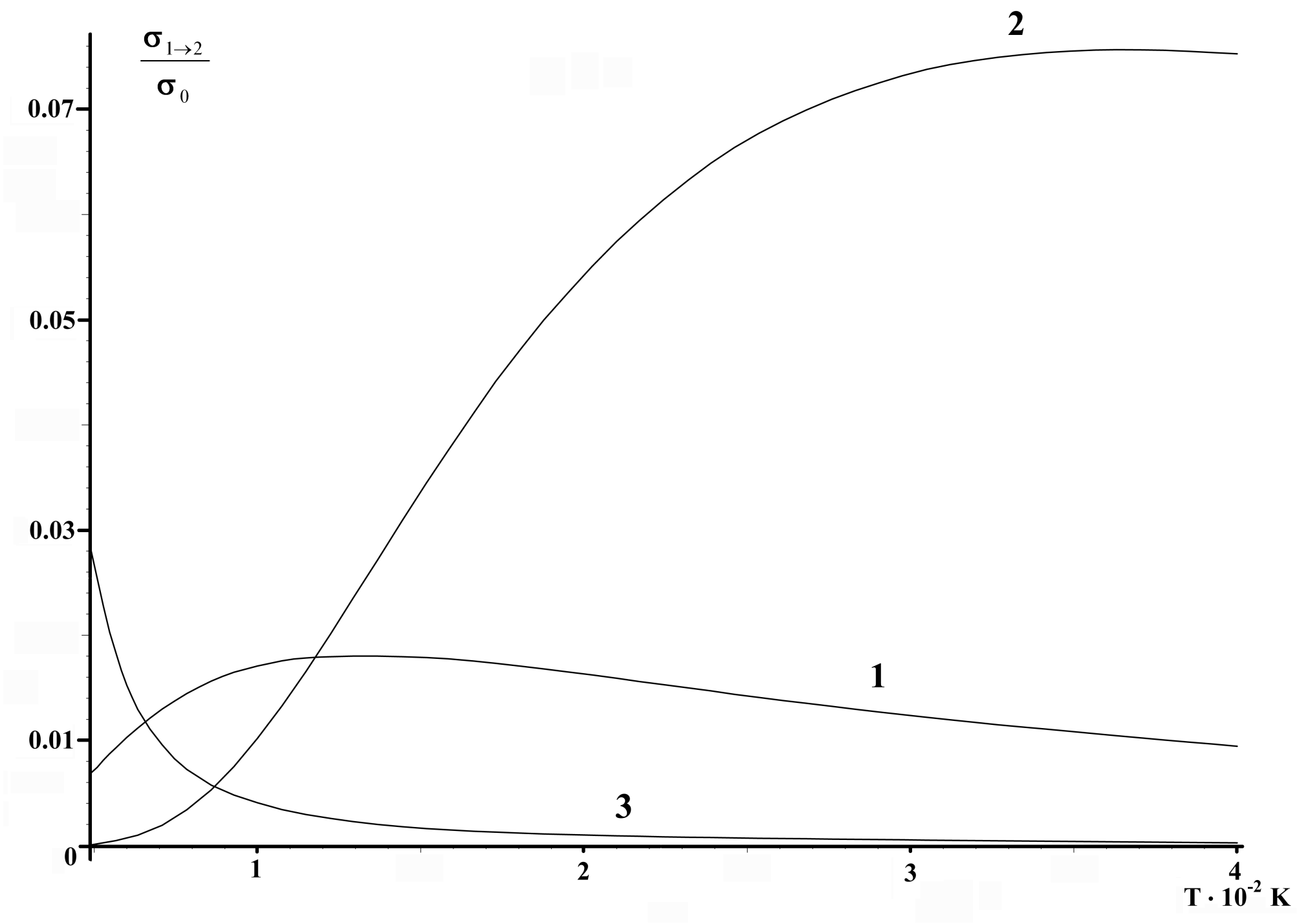

**Рис. 3**. Разложение скорости переноса дырки по вкладам от различных процессов: 1 – туннельный механизм, 2 – прыжковый механизм, 3 – обратный перенос, см. объяснения в тексте.

Теперь оценим значения параметров $A_1$ и $A_2$ в системе уравнений (28) – (31). При оценке будем ориентироваться на экспериментальные данные Льюиса с соавт. [37–41]. Как уже сказано во введении, в их работах исследовались олигомеры вида $St–X_n–G–Y_m–dG–dG–A_k$ или $St–X_n–G–Y_m–dG–dG–dG–A_k$, где $X$ и $Y$ – $dA$ или $dT$, $2 \le n \le 3$; $1 \le m \le 2$; $0 \le k \le 2$. Выбрав фрагмент с $n = 0$, и рассматривая только перенос дырки с первого $dG$ на ближайший с 5'-конца $dA$, перепишем систему (28) – (31) после усреднения по быстрым переменным в следующем виде:

$$\begin{aligned} &(\frac{d}{dt} + \alpha + k_{12})G_{11} - k_{12}G_{22} = 0, \\ &(\frac{d}{dt} + k_{12})G_{22} - k_{12}G_{11} = 0 \end{aligned} \qquad (36)$$

Её аналитическое решение имеет вид

$$G_{11} = C_1 \exp\left(\left(-k_{12} - \frac{\alpha}{2} + \sqrt{k_{12}^2 + \frac{\alpha^2}{4}}\right)t\right) + C_2 \exp\left(\left(-k_{12} - \frac{\alpha}{2} - \sqrt{k_{12}^2 + \frac{\alpha^2}{4}}\right)t\right),$$

или, при начальных условиях $G_{11}(0) = 1$, $\frac{d}{dt}G_{11}(0) = -k_{12} - \alpha$,

$$\begin{aligned} G_{11} = &\frac{1}{2}\left(1 - \frac{\alpha}{\sqrt{4k_{12}^2 + \alpha^2}}\right)\exp\left(\left(-k_{12} - \frac{\alpha}{2} + \sqrt{k_{12}^2 + \frac{\alpha^2}{4}}\right)t\right) + \\ &+ \frac{1}{2}\left(1 + \frac{\alpha}{\sqrt{4k_{12}^2 + \alpha^2}}\right)\exp\left(\left(-k_{12} - \frac{\alpha}{2} - \sqrt{k_{12}^2 + \frac{\alpha^2}{4}}\right)t\right) \end{aligned} \qquad (37)$$

Из выражения (37) следует, что заселенность сайта состоит из двух экспонент с разными параметрами затухания. Отсюда имеем две скорости релаксации – высокую

$$v_r = k_{12} + \frac{\alpha}{2} + \sqrt{k_{12}^2 + \frac{\alpha^2}{4}},$$

и низкую

$$v' = k_{12} + \frac{\alpha}{2} - \sqrt{k_{12}^2 + \frac{\alpha^2}{4}} \approx \frac{\alpha}{2} = 0,005,$$

Последнее равенство следует из предположения о гарпунном механизме рекомбинации ион-радикальных пар. Для данного механизма скорость рекомбинации пропорциональна квадрату резонансного интеграла для расстояний между $St^{\bullet-}$ и $dG^{\bullet+}$. Другими словами, время рекомбинации, $\tau_a$, пропорционально $L_0^{-2}e^{n\rho}$, где $L_0$ – резонансный интеграл между $St^{\bullet-}$ и $dG^{\bullet+}$ при $n = 0$, а $\rho$ – параметр, пропорциональный расстоянию между катион- и анион-радикалом.

Эти параметры можно определить из экспериментальных времён рекомбинации для случаев $n = 1$ ($\tau_a = 0{,}10$ нс) [37, 41], $n = 2$ ($\tau_a = 1{,}5$ нс) [37, 38, 39] и $n = 3$ ($\tau_a$ составляет 29 – 55 нс) [39, 41]. Сравнение рассчитанных нами времён с экспериментом приведено в Таблице 2. Найденные значения безразмерных параметров $B_0$ и $\exp(\rho)$ равны, соответственно, 1,57 и 24. Далее мы вычислили значение $k_{12}$ по формуле (35), что сделало оценку $(A_1 - A_2)^2$ тривиальной задачей, поскольку $k_{12} = \tau_a^{-1}$. Как будет показано далее, $(A_1 - A_2)^2 = 25$ для фрагмента $dG_1$–$dA_2$–$dG_3$–$dG_4$.

**Таблица 2.** Времена рекомбинации, $\tau_a$, электронов и дырок во фрагментах ДНК

| | Расчётное $\tau_a$ | Экспериментальное $\tau_a$ |
|---|---|---|
| $\{St^{\bullet-}\cdot\cdot dG^{\bullet+}$–$(dA_2)\}$ | 0,007 | – |
| $\{St^{\bullet-}\cdots dA$–$dG^{\bullet+}$–$(dA_2)\}$ | 0,10 | 0,1 [37, 41] |
| $\{St^{\bullet-}\cdots dA$–$dA$–$dG^{\bullet+}$–$(dA_2)\}$ | 2,4 | 0,92 [38, 41];<br>1,5 [37, 38, 39] |
| $\{St^{\bullet-}\cdots dA$–$dA$–$dA$–$dG^{\bullet+}$–$(dA_2)\}$ | 57 | 29 [39]; 55 [41] |
| $\{St^{\bullet-}\cdot\cdot dA$–$dA$–$dA$–$dA$–$dG^{\bullet+}$–$(dA_2)\}$ | $\approx 10^3$ | – |
| $\{St^{\bullet-}\cdot\cdot dA$–$dA$–$dA$–$dA$–$dA$–$dG^{\bullet+}$–$(dA_2)\}$ | $\approx 2{,}5 \cdot 10^4$ | – |

Теперь можно исследовать перенос дырки в этом фрагменте более точно. Для этого необходимо решить систему

$$
\begin{aligned}
& i\frac{d}{dt}G_{11} = -B_{12}G_{21} + B_{12}G_{12} - i\alpha G_{11}, \qquad G_{11}(0) = 1 \\
& i\frac{d}{dt}G_{12} = E_{12}G_{12} - B_{12}G_{22} + B_{12}G_{11} + F_{12}(t)G_{12} \\
& i\frac{d}{dt}G_{21} = -E_{12}G_{21} - B_{12}G_{11} + B_{12}G_{22} + F_{21}(t)G_{21} \\
& i\frac{d}{dt}G_{22} = -B_{12}G_{12} - B_{23}G_{32} + B_{23}G_{23} + B_{12}G_{21} \\
& i\frac{d}{dt}G_{23} = E_{23}G_{23} - B_{23}G_{33} + B_{23}G_{22} + F_{23}(t)G_{23} \\
& i\frac{d}{dt}G_{32} = -E_{23}G_{32} - B_{23}G_{22} + B_{23}G_{33} + F_{32}(t)G_{32} \\
& i\frac{d}{dt}G_{33} = -B_{23}G_{23} - B_{34}G_{43} + B_{23}G_{32} + B_{34}G_{34} \\
& i\frac{d}{dt}G_{34} = E_{34}G_{34} - B_{34}G_{44} + B_{34}G_{33} + F_{34}(t)G_{34} \\
& i\frac{d}{dt}G_{43} = -E_{34}G_{43} - B_{34}G_{33} + B_{34}G_{44} + F_{43}(t)G_{43} \\
& i\frac{d}{dt}G_{44} = -B_{34}G_{34} + B_{34}G_{43}
\end{aligned}
\qquad (38)
$$

с нулевыми начальными условиями во всех узлах кроме первого. Система (38) описывает перенос заряда по цепочке, заселенность узлов и скорость переноса заряда. Здесь $E_{mn} = -\varepsilon_m + \omega A_m^2 + \varepsilon_n - \omega A_n^2 = -E_{nm}$, $F_{mn}(t) = F_{nm}(t) = \omega(A_m - A_n)^2[Ne^{-i\omega t} - (N+1)e^{i\omega t}]$, $\omega$ и $N$ – частоты колебаний и числа заполнения колебательных состояний в окрестности соседних узлов $m$ и $n$. Система приведена к безразмерному виду так же, как и в работе [42]. Характеристическая единица времени составляет $10^{-14}$ с. Значения величин $B_{12}$, $B_{23}$ и $B_{34}$ взяты из работы [49] и в безразмерных единицах составляют: $B_{GA}$ = 1,352; $B_{AG}$ = 0,744; $B_{GG}$ = 1,276. Для последовательностей из пяти узлов решали систему, аналогичную системе (38), где $B_{AA}$ = 0,455, см. [49].

На данном этапе вычислений мы выбрали произвольные безразмерные параметры взаимодействия дырок с колебаниями, с учётом условия $(A_1 - A_2)^2 = 25$. Параметрам были заданы значения: $A_1$ = 3, $A_2$ = 8, $A_3$ = 11, $A_4$ = 2. В дальнейшем они корректировались из сравнения функций $G_{11}(t)$ с экспериментальными данными [37–41]. Частота колебаний $\omega$ была равна 0,01 безразмерной единицы, что соответствует $10^{12}$ с$^{-1}$ – характерной частоте колебаний нуклеотидной пары вдоль комплементарных водородных связей [42]. При таком выборе параметров $A_n$ имеем: $E_{12}$ = 6,34; $E_{23}$ = –6,84; $E_{34}$ = 0,63; $F_{12}(t)$ = $F_{21}(t)$ = $25\omega\cdot[Ne^{-i\omega t} - (N+1)e^{i\omega t}]$; $F_{23}(t)$ = $F_{32}(t)$ = $9\omega\cdot[Ne^{-i\omega t} - (N+1)e^{i\omega t}]$; $F_{34}(t)$ = $F_{43}(t)$ = $90\omega\cdot[Ne^{-i\omega t} - (N+1)e^{i\omega t}]$.

На Рис. 4 приведено сравнение расчётных значений $G_{11}(t)$ при T = 300 K с экспериментом для фрагмента $dG_1$–$dA_2$–$dG_3$–$dG_4$ [37]. Согласие характеристических времён ухода дырки с первого сайта свидетельствует о правильности разработанного аналитического подхода. Следует отметить, что, как и для всякой обратной задачи, удовлетворительное сочетание параметров может быть не единственным, аналогично работе [42].

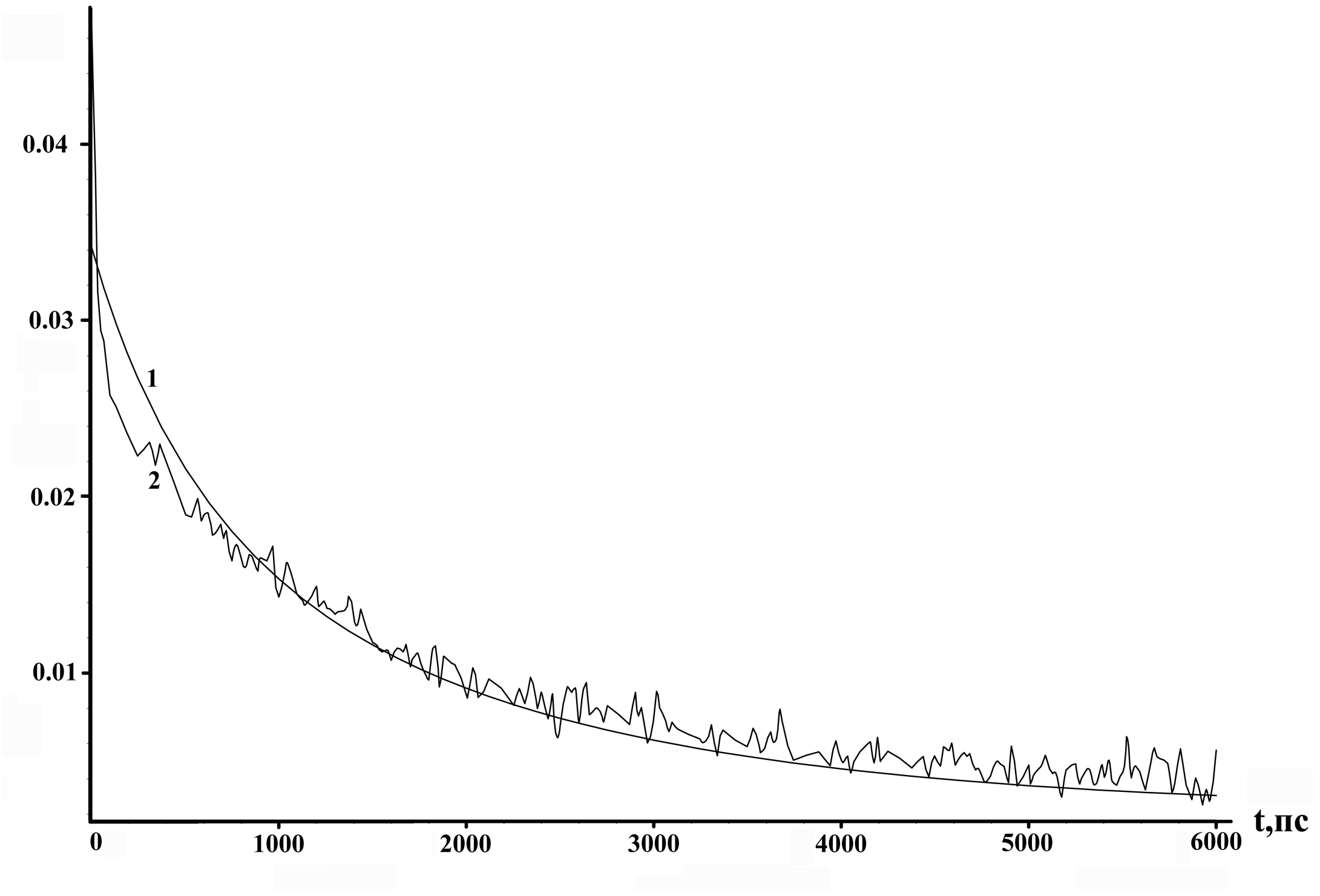

**Рис. 4**. Сравнение расчётной зависимости $G_{11}(t)$ (1) для фрагмента $dG_1$–$dA_2$–$dG_3$–$dG_4$ с экспериментально наблюдаемым кратковременным поглощением стильбенового анион-радикала пары $\{St^{\bullet -}\cdot\cdot dG^{\bullet +}\}$ в олигомере $St$–$dA$–$dA$–$dG_1$–$dA_2$–$dG_3$–$dG_4$ (2) [37].

Используя эти параметры, мы рассчитали корреляционные функции для всех узлов. Средние заселенности узлов с течением времени $G_{11}(t)$, $G_{22}(t)$, $G_{33}(t)$, $G_{44}(t)$, $G_{55}(t)$ были определены из кинетических уравнений, которые, в свою очередь, выведены из системы (38). Для этого мы предварительно усреднили по времени недиагональные члены. Аналогичные операции уже выполнялись при получении выражения (35), в случае малого фрагмента. Константы скоростей перехода дырки представлены в виде произведений $k_{mn} = B_{mn}^2 p_{mn}$, где $p_{mn}$ определяются по формуле $p_{mn} = \mathrm{Re}\int_0^\beta d\lambda \int_0^\infty dt \cdot e^{-st} < a_m^+(-i\lambda)a_n(-i\lambda)a_n^+(t)a_m(t) >$; $B_{mn}$ приведены выше, а значения $p_{mn}$ для ряда температур – в Таблице 3.

**Таблица 3.** Величины $p_{mn}$ при разных температурах

| | T = 200 K | T = 300 K | T = 400 K |
|---|---|---|---|
| $p_{GA}$ | 0,0066 | 0,0047 | 0,0037 |
| $p_{AG}$ | 0,0092 | 0,0058 | 0,0042 |
| $p_{GG}$ | 0,0185 | 0,0127 | 0,0095 |
| $p_{AA}$ | 0,0154 | 0,0118 | 0,0089 |

На Рис. 5, оформленном в виде таблицы графиков, показана динамика мгновенных и средних значений заселённости всех четырёх узлов последовательности $dG_1$–$dA_2$–$dG_3$–$dG_4$. В столбце А приведены заселённости $G_{11}(t)$, $G_{22}(t)$, $G_{33}(t)$, и $G_{44}(t)$ как функция времени при $\tau_a$ = 0,1 нс. В столбцах В и С вклад рекомбинации ИРП $\{St^{\bullet -}\cdot\cdot dG^{\bullet +}\}$ не учитывается ($\tau_a = \infty$), в целях наглядности.

| | **A**<br>Средняя заселённость как функция времени, для интервала 0 – 10000 пс, при $\tau_a$ = 0,1 нс и<br>T = 200, 300, 400K; | **B**<br>Мгновенная заселённость при T = 300 K для временного интервала 0 – 0,3 пс | **C**<br>Мгновенная заселённость при T = 300 K для временного интервала 2000 – 2000,3 пс |
|---|---|---|---|
| $G_{11}(t)$ | 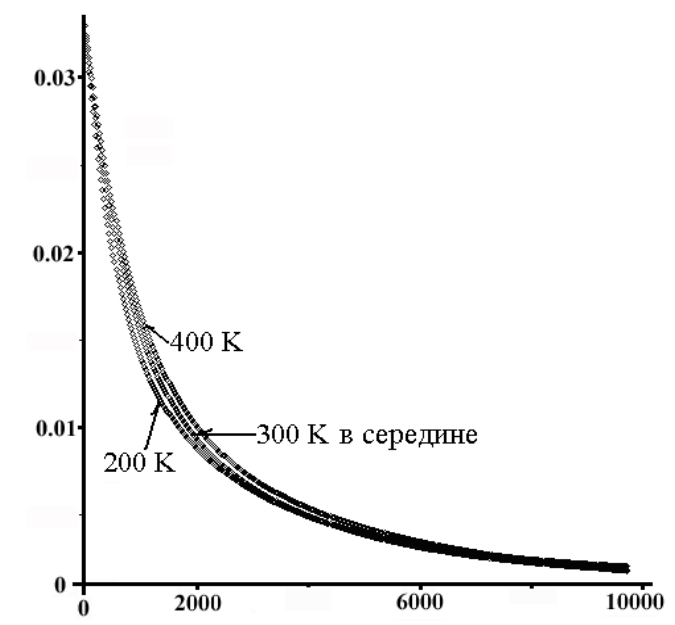 | 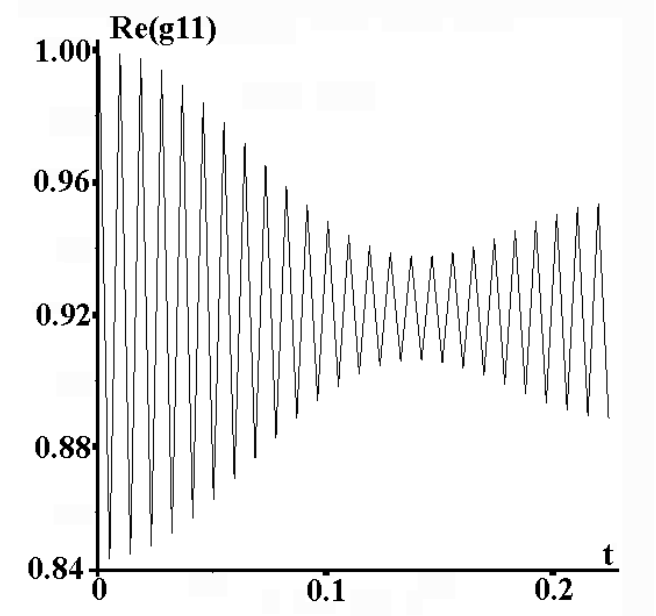 | 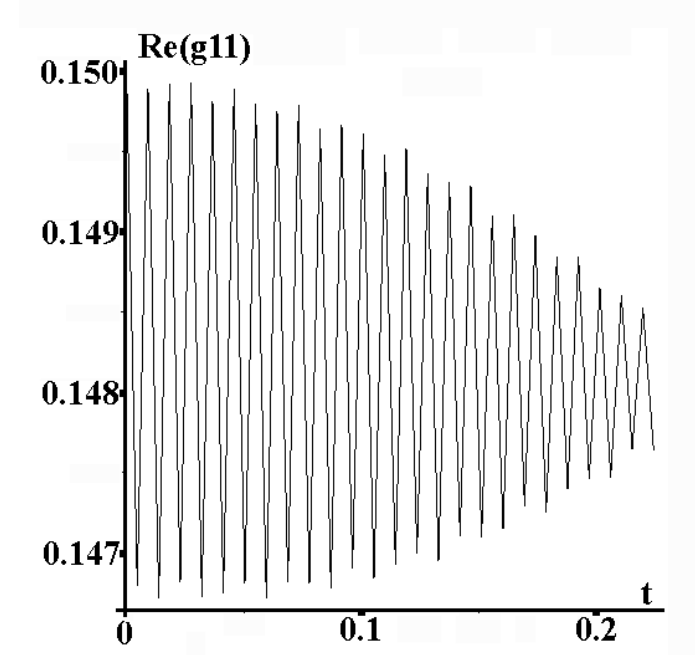 |
| $G_{22}(t)$ | 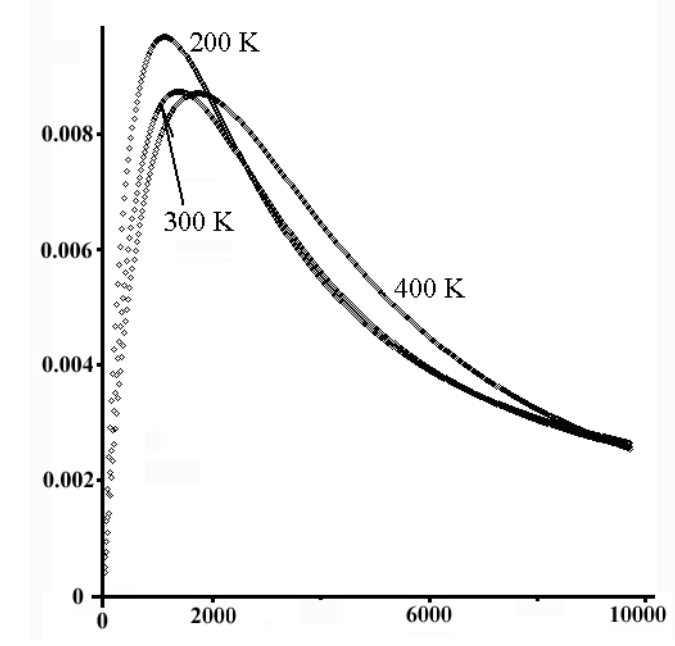 | 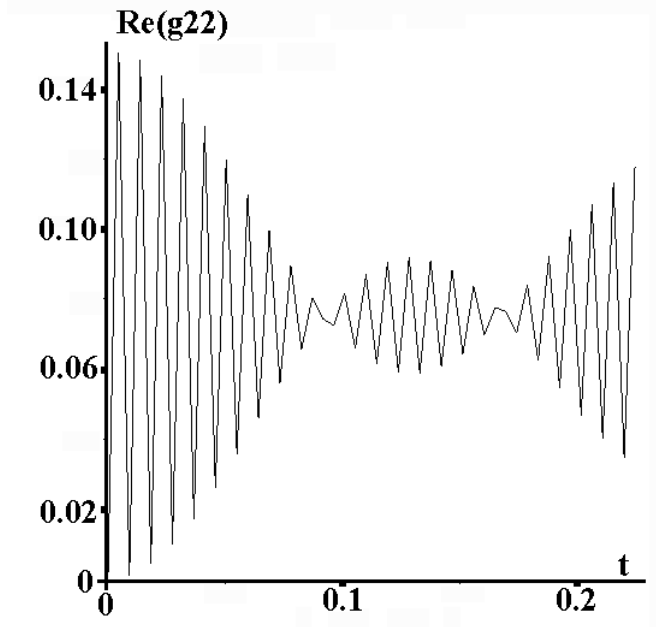 | 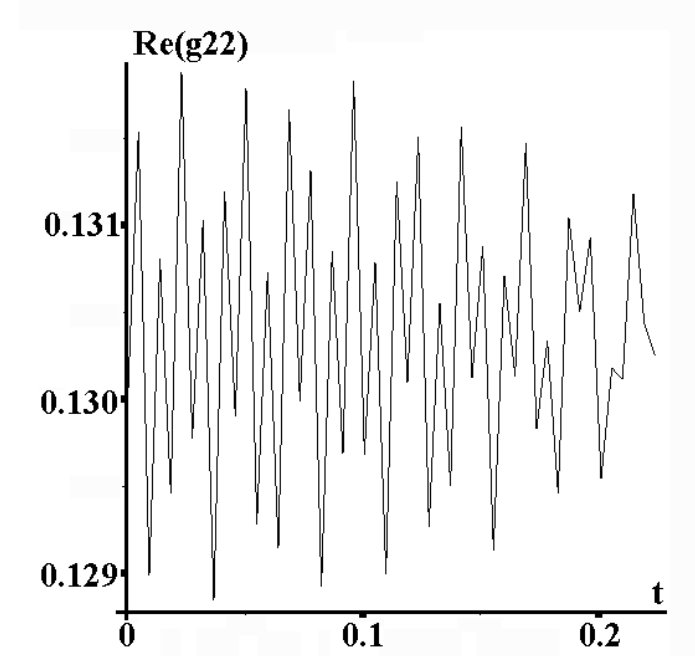 |
| $G_{33}(t)$ | 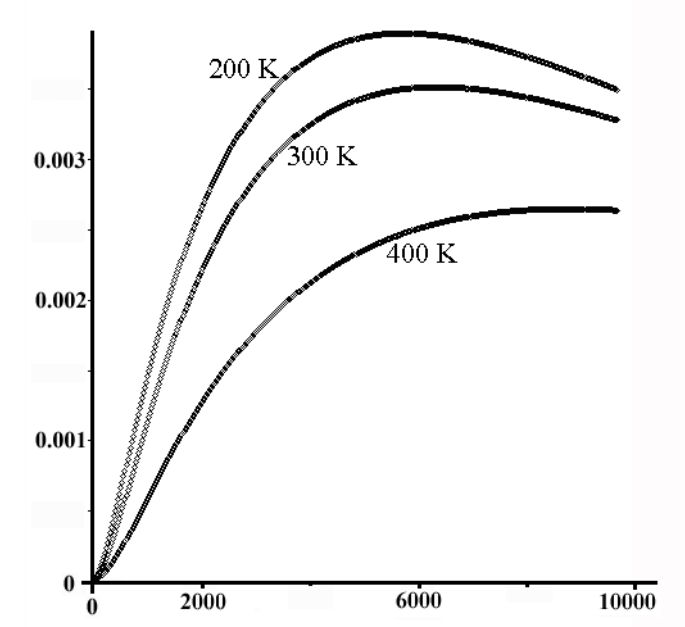 | Не заселены | 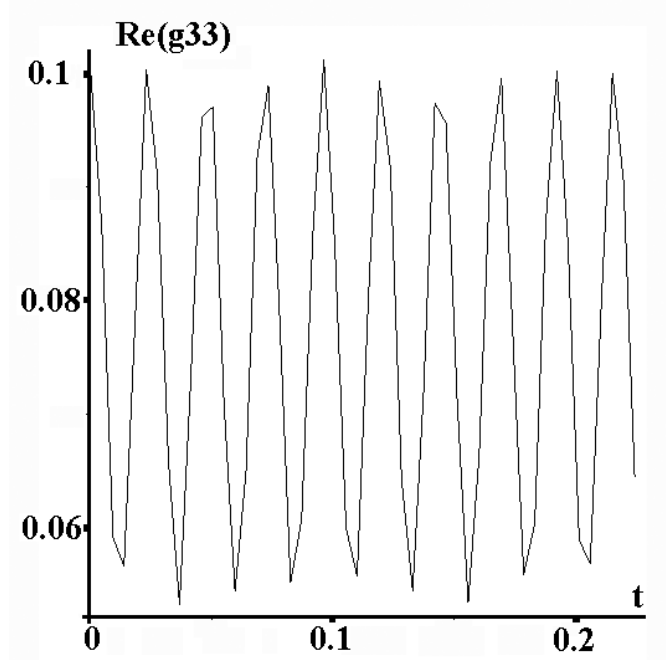 |
| $G_{44}(t)$ | 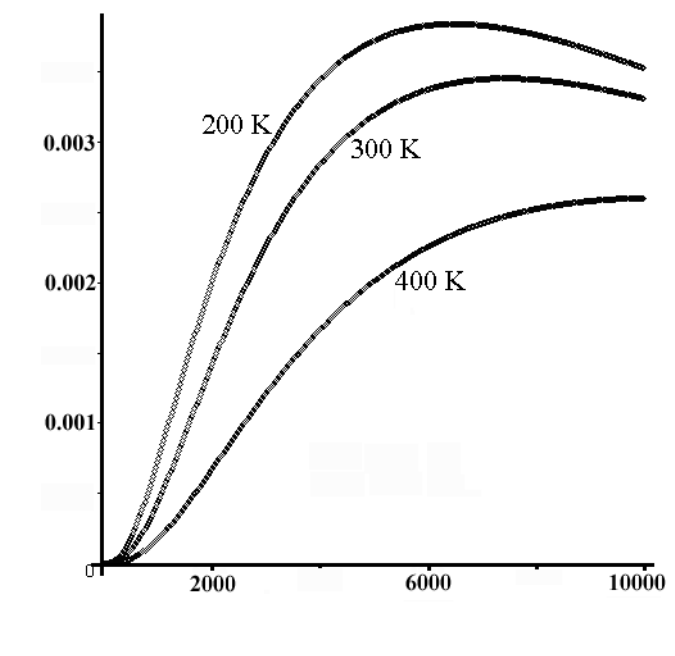 | Не заселены | 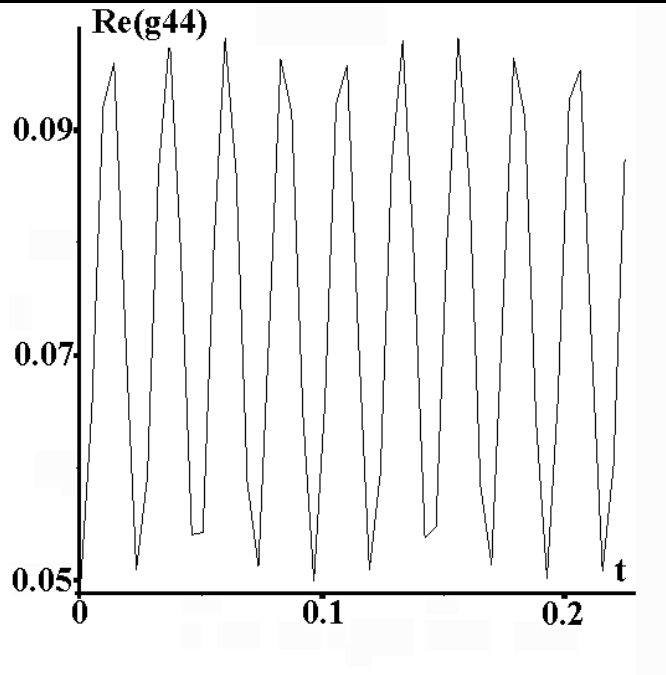 |

**Рис. 5**. Заселённости различных узлов последовательности $dG_1$–$dA_2$–$dG_3$–$dG_4$ как функция времени. Во всех графиках по оси абсцисс отложено время в пикосекундах. В столбце **A** представлены усреднённые по осцилляциям значения заселённости узлов. Данные заселённости показаны как функции времени (в пс) при температурах 200, 300 и 400 K. В столбцах **B** и **C** показаны мгновенные значения заселённости тех же узлов при T = 300 K во временном интервале 0 – 0,3 пс (**B**) и 2000 – 2000,3 пс (**C**).

Для мгновенной заселённости, помимо интервалов, отображённых на рисунках в столбцах B и C, исследован целый ряд других малых отрезков времени. Во всех случаях результаты получались одни и те же.

Сравнивая попарно картинки внутри столбцов B и C на Рис. 5, легко видеть, что колебания мгновенной заселённости первого и второго сайтов находятся в противофазе, как и колебания заселённости третьего и четвертого. Следовательно, быстрые квантовые осцилляции плотности заряда происходят в пикосекундном масштабе времени, что качественно соответствует более ранним расчётам, проведённым в нашей лаборатории [52]. Что интересно, быстрого обмена между вторым и третьим сайтами не наблюдалось. Дырка была локализована, в основном, либо в начале фрагмента, либо в его конце.

Путём сопоставления расчётных данных с результатами экспериментов [37–41], удалось подобрать параметры взаимодействия зарядов с колебаниями для нашей модели. Для $\omega = 0,01$ их величины составили: $A_1 = 3$, $A_2 = 8$, $A_3 = 11$, $A_4 = 2$. В Таблице 4 приведены основные результаты вычислений для семи фрагментов ДНК, исследованных Льюисом с соавт. [37–41], отражающие кинетику установления равновесия.

**Таблица 4.** Зависимость скорости релаксации дырки и заселенности сайтов от строения фрагмента ДНК; расчётные величины $\tau_a$ приведены для удобства. Нормировка заселенности первого сайта $G_{11}(0) = 1$, то есть рекомбинация скомпенсирована. Величина $\frac{\max}{ns}$ означает максимальное значение заселенности (числитель) и время, когда оно достигается, которое указывается в наносекундах (знаменатель)

| № | Фрагмент | $\tau_a$ , нс | $G_{22}(t)$ $\frac{\max}{ns}$ | $G_{33}(t)$ $\frac{\max}{ns}$ | $G_{44}(t)$ $\frac{\max}{ns}$ | $G_{55}(t)$ $\frac{\max}{ns}$ |
|---|---|---|---|---|---|---|
| 1 | $dG_1$–$dA_2$–$dG_3$–$dG_4$ | 0,007 | $\frac{0,4}{0,4}$ | $\frac{0,09}{4}$ | $\frac{0,19}{10}$ | – |
| 2 | $dA$–$dG_1$–$dA_2$–$dG_3$–$dG_4$ | 0,10 | $\frac{0,46}{0,8}$ | $\frac{0,19}{10}$ | $\frac{0,19}{10}$ | – |
| 3 | $dA$–$dA$–$dG_1$–$dA_2$–$dG_3$–$dG_4$ | 2,4 | $\frac{0,46}{0,8}$ | $\frac{0,23}{20}$ | $\frac{0,23}{20}$ | – |
| 4 | $dA$–$dA$–$dA$–$dG_1$–$dA_2$–$dG_3$–$dG_4$ | 57 | $\frac{0,46}{0,8}$ | $\frac{0,25}{20}$ | $\frac{0,25}{20}$ | – |
| 5 | $dA$–$dA$–$dG_1$–$dA_2$–$dG_3$–$dG_4$–$dG_5$ | 2,4 | $\frac{0,46}{0,8}$ | $\frac{0,18}{20}$ | $\frac{0,18}{20}$ | $\frac{0,19}{20}$ |
| 6 | $dA$–$dA$–$dA$–$dG_1$–$dA_2$–$dA_3$–$dG_4$–$dG_5$ | 57 | $\frac{0,50}{0,8}$ | $\frac{0,18}{100}$ | $\frac{0,18}{100}$ | $\frac{0,19}{100}$ |
| 7 | $dA$–$dA$–$dG_1$–$dA_2$–$dA_3$–$dG_4$–$dG_5$ | 2,4 | $\frac{0,50}{0,8}$ | $\frac{0,18}{100}$ | $\frac{0,18}{100}$ | $\frac{0,19}{100}$ |

В первом фрагменте, для которого $\tau_a$ очень мало, функции $G_{33}(t)$ и $G_{44}(t)$ достигают своих максимумов наиболее быстро. Система становится близка к равновесию уже через 5 нс после образования ион-радикальной пары. Изменения соотношения заселённостей на более поздних временах довольно незначительны.

Для второго фрагмента время рекомбинации в 14 раз больше, скорость достижения функциями $G_{33}(t)$ и $G_{44}(t)$ своих максимумов несколько ниже, и даже через 5 нс после разделения зарядов система ещё далека от равновесия. Характерное $\tau_a$ пары $\{St^{\bullet-} \cdot\cdot dG^{\bullet+}\}$ в третьем фрагменте находится уже в масштабе наносекунд, превышая аналогичное время для второго фрагмента в 24 раза. Вследствие этого, времена достижения функциями $G_{33}(t)$ и $G_{44}(t)$ своих максимумов возрастают ещё в два раза, хотя для $G_{22}(t)$ этот показатель уже не изменяется.

Таким образом, скорость рекомбинации ион-радикальной пары является важным фактором, определяющим время установления равновесия при переносе заряда в гетерогенных олигонуклеотидах. Сравнив данные для третьего и четвёртого фрагментов, можно видеть, что, несмотря на разные $\tau_a$, динамика их функций $G_{22}(t)$, $G_{33}(t)$ и $G_{44}(t)$ фактически одинакова. Из этого можно заключить, что, начиная с некоторого своего значения, время рекомбинации ИРП $\{St^{\bullet -} \cdots dG^{\bullet +}\}$ перестаёт влиять на скорость установления равновесного распределения заряда.

Особенностью пятого фрагмента является наличие на его 5'-конце трёх гуаниновых оснований вместо двух. Единственным следствием этой особенности являются несколько меньшие величины максимумов $G_{33}(t)$, $G_{44}(t)$ и $G_{55}(t)$, по сравнению с максимумами $G_{33}(t)$ и $G_{44}(t)$ в четвёртом фрагменте. Анализ динамики установления равновесия показал, что на большей части временного интервала сумма $G_{33}(t) + G_{44}(t)$ четвертого фрагмента очень близка к $G_{33}(t) + G_{44}(t) + G_{55}(t)$ пятого.

В шестом и седьмом фрагментах гуаниновые основания разделены не одним *dA*, а двумя. Из-за этого времена достижения функциями $G_{33}(t)$, $G_{44}(t)$ и $G_{55}(t)$ максимумов увеличиваются до 100 нс, вне зависимости от $\tau_a$. Снижение скорости переноса в 5 раз при увеличении промежутка между *dG* на одно *dA* является результатом, достаточно близким к экспериментальным данным, согласно которым аналогичная величина составляет 6–10 раз [29, 33, 40].

Что интересно, величины максимального значения $G_{33}(t)$ этих фрагментов примерно те же, что и для пятого, хотя у последнего третьим узлом является не *dA*, а *dG*. Для выяснения физической природы этой особенности, а также ряда других характерных черт поведения разработанной модели требуются дальнейшие исследования, которые проводятся нами в настоящее время.

## ЗАКЛЮЧЕНИЕ

Полученные результаты демонстрируют важные физические закономерности переноса катион-радикала в гетерогенной ДНК. Согласно нашим расчетным данным, миграция заряда является процессом, протекающем обычно в наносекундном масштабе времени, что хорошо согласуется с экспериментами [29, 32, 33, 37–41]. Тем не менее, в переносе заряда имеют место квантовые осцилляции плотности его вероятности, период которых измеряется долями пикосекунды. Этот результат, в свою очередь, качественно совпадает с расчётными данными, полученными в нашей лаборатории ранее [52].

В работе показано, что скорость перехода заряда между соседними основаниями почти не зависит от динамики более удалённых нуклеотидных пар. В процессе переноса большую роль играет эффективная поляризация, обусловленная взаимодействием дырок с тепловыми колебаниями нуклеотидных пар. Величина этих взаимодействий может превышать энергию колебаний оснований в 5–10 раз.

Разработанный нами подход может быть в дальнейшем улучшен и адаптирован для изучения более сложных гамильтонианов. В частности, в нём можно учесть произвольную зависимость матричных элементов перехода от расстояний между сайтами (эффект Кондона), а также изменение частот колебаний при переходе с сайта на сайт (эффект Душинского и частотный эффект) [53]. По всей видимости, данный подход окажется эффективным и при рассмотрении различных нелинейных моделей – например, модели Якушевич [54], Пейярда-Бишопа [48], Пейярда-Бишопа-Доксуа [55], моделей с ангармонизмом Буссинеска и других гамильтонианов, см. [53].

Изложенный метод получения замкнутых уравнений корреляционных функций пригоден для произвольных систем, в том числе и для ДНК. Важно, что он имеет внутренний критерий применимости: достаточно лишь провести вычисления с учётом корреляционных функций более высокого порядка и, сопоставив полученные результаты, определить область их сходимости.

Физический смысл функций $M_q^+(t)$, $M_q^-(t)$, $D_{q_1j_1;q_2j_2}^{+,-}(t)$, $D_{q_1j_1;q_2j_2}^{-,+}(t)$, $D_{q_1j_1;q_2j_2}^{+,+}(t)$, $D_{q_1j_1;q_2j_2}^{-,-}(t)$ и т.д. состоит во введении переменных самосогласованных полей вместо операторов под знаком усреднения. Это существенно упрощает как аналитические расчёты, так и вычислительные эксперименты. Возможности предлагаемого метода исследованы недостаточно, однако известно, что он хорошо учитывает эффекты ближнего порядка и хорошо аппроксимирует КФ на малых временах. Метод открывает новые возможности, связанные с появлением нелинейных уравнений. Более того, он позволяет построить нестандартную теорию возмущений, которая уже в нулевом порядке учитывает схождение уровней к границе диссоциации [53]. Это свойство открывает большие перспективы в исследовании самых разнообразных систем. Аналогичное свойство, например, нелинейной модели Пейярда-Бишопа-Доксуа, сделало её мощнейшим инструментом исследования динамики локальных денатурированных областей ДНК при T < 310 K [56, 57].

Область применения полученных результатов может быть чрезвычайно широка. Помимо изучения закономерностей переноса катион-радикалов в ДНК живой клетки, исследования в этой области крайне важны для разработки нанобиоэлектронных устройств на основе ДНК. Получение надёжного способа эффективного, быстрого и точного расчёта проводимости ДНК любой длины и нуклеотидной последовательности является одной из ключевых задач в этой области.

Кроме того, разработанный нами метод можно использовать для решения многих нелинейных задач, в том числе и не имеющих отношения к ДНК.

Авторы благодарны Российскому Фонду фундаментальных исследований, поддержавшему эту работу (гранты №12-07-00279, №12-07-33006, №11-07-00635, №11-07-12054).

## СПИСОК ЛИТЕРАТУРЫ